\documentclass[preprint,12pt,3p,authoryear]{aastex631}

\definecolor{mygray}{gray}{0.9}

\usepackage{graphicx}
\usepackage[colorlinks=true]{hyperref}
\usepackage[utf8]{inputenc}
\usepackage{amsmath} 
\usepackage{xcolor}
\usepackage{threeparttable}

\usepackage{lipsum}
\usepackage{mathtools}
\usepackage{cuted}

\begin{document}


\newcommand{\tnm}[1]{{\tablenotemark{#1}}}
\newcommand{\tnt}[2]{{\tablenotetext{#1}{#2}}}
\newcommand{\tab}[1]{{table\,\ref{#1}}}
\newcommand{\eqn}[1]{{eq.\,\ref{#1}}}

\newcommand{\Hawaii}{{Hawai`i}}

\newcommand{\eg}{{\it e.g.}}
\newcommand{\ie}{{\it i.e.}}
\newcommand{\etc}{{\it etc.}}
\newcommand{\etal}{{\it et~al.}}
\newcommand{\adhoc}{{\it ad~hoc}}
\newcommand{\insitu}{{\it in situ}}
\newcommand{\apriori}{{\it a~priori}}
\newcommand{\postfacto}{{\it post~facto}}

\newcommand{\half}{{\frac{1}{2}}}
\newcommand{\mean}[1]{\langle{#1}\rangle}
\newcommand{\dif}{\mathrm{d}}
\newcommand{\overbar}[1]{\mkern 1.5mu\overline{\mkern-1.5mu#1\mkern-1.5mu}\mkern 1.5mu}
\newcommand{\oforder}{\mathcal{O}}
\newcommand{\sci}[2]{{{#1}\times10^{#2}}}

\newcommand{\xbar}{\bar x}
\newcommand{\xyz}{(x,y,z)}
\newcommand{\xyzdot}{(\dot{x},\dot{y},\dot{z})}
\newcommand{\aei}{(a,e,i)}
\newcommand{\qei}{(q,e,i)}
\newcommand{\aeiH}{(a,e,i,H)}
\newcommand{\qeiD}{(q,e,i,D)}
\newcommand{\OoM}{(\Omega,\omega,M)}
\newcommand{\vo}{\vec{o}}
\newcommand{\vx}{\vec{x}}
\newcommand{\vxdot}{\vec{\dot x}}
\newcommand{\vxavg}{\mean{\vec{x}}}
\newcommand{\vy}{\vec{y}}
\newcommand{\vyavg}{\mean{\vec{y}}}
\newcommand{\vz}{\vec{z}}
\newcommand{\vzavg}{\mean{\vec{z}}}
\newcommand{\Havg}{\mean{H}}
\newcommand{\deltav}{\Delta v}
\newcommand{\node}{\Omega}
\newcommand{\aperi}{\omega}
\newcommand{\lperi}{\tilde\omega}

\newcommand{\todo}[2]{{\color{red}\bf #1 - #2}}
\newcommand{\note}[1]{{\color{blue}{\bf #1}}}
\newcommand{\XXX}{{\color{red}\bf XXX}}
\newcommand{\citepna}[1]{{{\color{red} (#1)}}}
\newcommand{\citetna}[1]{{{\color{red} #1}}}

\newcommand{\sn}{$S/N$}
\newcommand{\SN}{$S/N$}
\newcommand{\rd}{$^{rd}$}
\newcommand{\st}{$^{st}$}
\newcommand{\nd}{$^{nd}$}
\newcommand{\fManx}{{\mathrm{f}_\mathrm{Manx}}}
\newcommand{\digesttwo}{{\texttt{digest2}}}

\newcommand{\HtwoO}{{H$_2$O}}
\newcommand{\CO}{{CO}}
\newcommand{\COtwo}{{CO$_2$}}

\newcommand{\h}{^\mathrm{h}}
\newcommand{\m}{^\mathrm{m}}
\newcommand{\s}{^\mathrm{s}}
\newcommand{\Mpc}{\,\mathrm{Mpc}}
\newcommand{\kpc}{\,\mathrm{kpc}}
\newcommand{\pc}{\,\mathrm{pc}}
\newcommand{\au}{\,\mathrm{au}}
\newcommand{\km}{\,\mathrm{km}}
\newcommand{\kph}{\,\mathrm{km}/\mathrm{h}}
\newcommand{\kps}{\,\mathrm{km}\,\mathrm{s}^{-1}}
\newcommand{\ft}{\,\mathrm{ft}}
\newcommand{\meter}{\,\mathrm{m}}
\newcommand{\cm}{\,\mathrm{cm}}
\newcommand{\mm}{\,\mathrm{mm}}
\newcommand{\um}{\,\mu \mathrm{m}}
\newcommand{\nm}{\,\mathrm{nm}}
\newcommand{\rad}{\,\mathrm{rad}}
\newcommand{\rms}{\,\mathrm{(rms)}}
\newcommand{\anno}{\,\mathrm{a}}
\newcommand{\yr}{\,\mathrm{yr}}
\newcommand{\Myr}{\,\mathrm{Myr}}
\newcommand{\Gyr}{\,\mathrm{Gyr}}
\newcommand{\Day}{\,\mathrm{day}}
\newcommand{\days}{\,\mathrm{d}}
\newcommand{\dayperyear}{\,\mathrm{d}/\mathrm{yr}}
\newcommand{\vrk}{\,\mathrm{vrk}}
\newcommand{\degrees}{\,\mathrm{deg}}
\newcommand{\hours}{\,hours}
\newcommand{\hour}{\,\mathrm{h}}
\newcommand{\hourperday}{\,\mathrm{h}/\mathrm{d}}
\newcommand{\minute}{\,\mathrm{min}}
\newcommand{\second}{\,\mathrm{s}}
\newcommand{\mps}{\,\meter\,\second^{-1}}
\newcommand{\Hz}{\,\mathrm{Hz}}
\newcommand{\mags}{\,\mathrm{mag}}
\newcommand{\K}{\,\mathrm{K}}
\newcommand{\J}{\,\mathrm{J}}
\newcommand{\N}{\,\mathrm{N}}
\newcommand{\kg}{\,\mathrm{kg}}
\newcommand{\g}{\,\mathrm{g}}
\newcommand{\AMU}{\,\mathrm{AMU}}
\newcommand{\W}{\,\mathrm{W}}
\newcommand{\MW}{\,\mathrm{MW}}
\newcommand{\degC}{\arcdeg\mathrm{C}}
\newcommand{\degK}{\mathrm{K}}
\newcommand{\Jy}{\,\mathrm{Jy}}
\newcommand{\mJy}{\,\mathrm{mJy}}
\newcommand{\Mearth}{\,\mathrm{M}_\oplus}

\newcommand{\asteroid}[2]{{({#1})\,{#2}}}
\newcommand{\designation}[2]{{{#1}\,{#2}}}
\newcommand{\TC}{{2008\,TC$_3$}}
\newcommand{\RH}{{2006\,RH$_{120}$}}
\newcommand{\Bennu}{{(101955)\,Bennu}}

\newcommand{\Sthree}{{C/2014$\,$S3$\,$PANSTARRS}}
\newcommand{\Uone}{{1I/2017 U1 (‘Oumuamua)}}


\newcommand{\gps}{\ensuremath{g_{\rm P1}}}
\newcommand{\rps}{\ensuremath{r_{\rm P1}}}
\newcommand{\ips}{\ensuremath{i_{\rm P1}}}
\newcommand{\zps}{\ensuremath{z_{\rm P1}}}
\newcommand{\yps}{\ensuremath{y_{\rm P1}}}
\newcommand{\wps}{\ensuremath{w_{\rm P1}}}
\newcommand{\grizy}{\gps\rps\ips\zps\yps}
\newcommand{\JHK}{\ensuremath{JHK}}
\newcommand{\V}{\ensuremath{V}}

\newcommand{\PS}{\protect \hbox {Pan-STARRS}}
\newcommand{\PSone}{\protect \hbox {Pan-STARRS1}}
\newcommand{\PStwo}{\protect \hbox {Pan-STARRS2}}
\newcommand{\PSfour}{\protect \hbox {Pan-STARRS4}}
\newcommand{\knownserver}{{\tt known\_server}}


\title{Total Solar Eclipse White Light Images as a Benchmark for PFSS Coronal Magnetic Field Models: An In-Depth Analysis over a Solar Cycle}

\author{Luke Fushimi Benavitz}
\affil{Institute for Astronomy, University of Hawai`i, Honolulu, HI 96822, USA}
\author{Benjamin Boe}
\affil{Wentworth Institute of Technology, Boston, MA 02115, USA}
\affil{Institute for Astronomy, University of Hawai`i, Honolulu, HI 96822, USA}
\author{Shadia Rifai Habbal}
\affil{Institute for Astronomy, University of Hawai`i, Honolulu, HI 96822, USA}

\begin{abstract}

Potential Field Source Surface (PFSS) models are widely used to simulate coronal magnetic fields. PFSS models use the observed photospheric magnetic field as the inner boundary condition and assume a perfectly radial field beyond a ``Source Surface" ($R_{ss}$). At present, total solar eclipse (TSE) white light images are the only data that delineate the coronal magnetic field from the photosphere out to several solar radii ($R_\odot$). We utilize a complete solar cycle span of these images between 2008 and 2020 as a benchmark to assess the reliability of PFSS models. For a quantitative assessment, we apply a rolling Hough transform (RHT) to the eclipse data and corresponding PFFS models to measure the difference, $\Delta\theta$, between the data and model magnetic field lines throughout the corona. We find that the average $\Delta\theta$, $\langle\Delta\theta\rangle$, can be minimized for a given choice of $R_{ss}$ depending on the phase within a solar cycle. In particular, $R_{ss}\approx1.3 \ R_\odot$ is found to be optimal for solar maximum,  while $R_{ss}\approx3 \ R_\odot$ yields a better match at solar minimum. However, large ($\langle\Delta\theta\rangle>10^\circ$) discrepancies between TSE data and PFSS-generated coronal field lines remain regardless of the choice of source surface. Yet, implementation of solar cycle dependent $R_{ss}$ optimal values do yield more reliable PFSS-generated coronal field lines for use in models and for tracing in-situ measurements back to their sources at the Sun.

\end{abstract}



\section{Introduction}
\label{s.Introduction}

The shape of the solar corona as defined by total solar eclipse (TSE) white light images, originally recorded through hand-drawn sketches and early photographic techniques \citep{Maunder1988}, was found to change with the sunspot cycle \citep{Darwin1899, Schwabe1844}. When the magnetic nature of sunspots was discovered by \cite{Hale1908} using the Zeeman effect, it became evident that the variability of the shape of the corona was driven by magnetic fields emerging from the photosphere and expanding into space. 

During the approximate one year duration of any solar minimum, the shape of the corona is invariably dominated by large polar plumes \citep{Saito1958} in the north and south poles \citep{Munro1972} with large streamers confined mostly to equatorial regions. In contrast,  coronal structures at solar maximum become much more complex with streamers appearing at all latitudes  \citep{Newkirk1967}. 
The close connection between the shape of the corona and the magnetic cycle was the first direct evidence that fine scale structures in the corona are indeed shaped by magnetic field lines. TSE white light images thus become the best proxy for inferring the direction of coronal magnetic fields starting from the solar limb out to at least 6 solar radii ($R_\odot$) (\citealt{Boe2020}).

One of the first and most widely used models of global coronal magnetic field, known as the Potential Field Source Surface (PFSS) model, was developed by \cite{Altschuler1969} and \cite{Schatten1969}. 
These authors extrapolated the photospheric magnetic field outward to an upper boundary, a heliocentric radius known as the “Radial Source Surface” ($R_{ss}$), beyond which the coronal magnetic field was assumed to become radial. Driven by TSE white light images at solar minimum, the original convention was to set at $R_{ss} = 2.5 ~ R_\odot$. With PFSS simulations, the corona is assumed to be current-free. This assumption is reasonable for regions with low plasma $\beta$ (i.e., ratio of plasma to magnetic pressure). However, the assumption may break down when the thermal pressure becomes comparable to the magnetic pressure (see Section \ref{s.Summary and Conclusions}).

Subsequently, 
\citet{Schulz1978} and \citet{Sakurai1979,Sakurai1981} explored the impact of  a non-spherical and varying $R_{ss}$. Sakurai called the varying force-free parameter $\alpha$; the ratio of electric current to the magnetic field strength. \citet{Schulz1978} found that a non-spherical source surface was a better match to magnetohydrodynamic (MHD) models, and the source surface was not necessarily convex.
\citet{Levine1982} implemented a range of $\alpha$ values in their MHD model and used the 1973 TSE white light image for comparison. They
found that a non-spherical $R_{ss}$ matched the white light image better; however,
the area and placement of the coronal holes did not match. \cite{Badman2020} used the PFSS approach to trace in-situ measured magnetic fields back to a Solar
Dynamic Observatory's Atmospheric Imaging Assembly (SDO/AIA) synoptic map of the corona. They concluded that $R_{ss}$ values between $1.3 - 1.5 \ R_\odot$ 
led to a better `landing' in `open' field regions commonly associated with coronal holes. \cite{Asvestari2019} compared a combination of the PFSS and Schatten Current Sheet (SCS) model to extreme ultraviolot (EUV) observations and Collection of Analysis Tools for Coronal Holes (CATCH). They found that common `default' heights of $2.3 - 2.6 \ R_\odot$ fail to accurately model coronal hole areas, and $R_{ss}$ values below $2.3 \ R_\odot$ improve the model.

Coronal magnetic fields remain  essential boundary conditions for MHD models of coronal heating and solar wind acceleration (e.g. MASA: \citealt{Mikic1999,Linker1999}; AWSoM: \citealt{vanDerHolst2014}). They are also essential for studying the formation and evolution of shocks driven by the expansion of coronal mass ejections \citep{Maguire2020}, the properties of the solar wind, and identifying its potential sources at the Sun \citep{Zhao2017, Song2023}. They are also critical for establishing the connectivity between in-situ magnetic field measurements and their sources at the Sun, as 
recently applied to the in-situ magnetic field measurements from the Parker Space Probe \citep{Badman2020}. Hence, assessing the reliability of Potential Field Source Surface (PFSS) models in producing global coronal magnetic fields in the continued absence of their direct measurement remains critical.

At present, the unsurpassed spatial resolution of TSE white light images, spanning at least 10 $  R_\odot$ above the solar surface, offers the only reliable visual rendition of coronal magnetic structures \citep{Habbal2021} to which the output of PFSS models can be compared. Recently, \citet{Boe2020} used TSE white light images acquired between 2001 and 2019 to measure the coronal magnetic field angle direction with respect to radial, by  applying the rolling Hugh transform (RHT) \citep{Clark2014} to white light TSE images. They found that field lines become radial between 4 and 5 $ \ R_\odot$. 

The goal of this work is to assess the reliability of PFSS models (Section \ref{s.pfss}) in generating a realistic coronal magnetic field by using white light total solar eclipse images as a benchmark (Sections ~\ref{s.data}, \ref{ss.The PFSS Model}). The RHT process provides a means for quantitative comparisons between the two (Section \ref{ss.RHT}). 
The limitations of the PFSS approach compared to the eclipse data are discussed in Sections \ref{s.Discussion} and \ref{s.Summary and Conclusions}. One of the outcomes of this comparison was the finding that more reliable PFSS models can be achieved with the implementation of a solar cycle dependent $R_{ss}$ (Section \ref{ss.Solar Cycle Effects}) (similar to \citealt{Lee2011, Arden2014}).



\section{TSE Data} 
\label{s.data}

Produced by Thompson scattering of the photospheric radiation by coronal electrons, processed TSE white light images \citep{{Druckmuller2006,Druckmuller2009}} yield the highest spatial resolution images, available at present, of the traces of coronal magnetic field lines out to at least 6 $ R_\odot$ above the limb. The TSE white light images used as a benchmark in this study were acquired from observations covering a solar cycle between 2008 and 2020 (see \citealt{Boe2020, Boe2021b, Habbal2021}). The TSE white light images are preferentially tracing the field lines in the plane-of-sky mainly due to Thompson scattering \citep{Howard2009}. This leads to off-axis, out of plane, emission being small compared to emission in the plane-of-sky (see Section~\ref{s.Discussion}, Appendix~\ref{s.Appendix}).

\section{Methodology for PFSS-Generated Coronal Magnetic Field Lines}
\label{s.pfss}

PFSS models of the global coronal magnetic field generally use synoptic maps of the photospheric magnetic field as an inner boundary condition. These maps require a full solar rotation of observations to include all longitudes. Thus, one of the main limitations of synoptic maps is the assumption that the corona does not change over a solar rotation. To account for these changes, we use the National Solar Observatory (NSO) Air Force Data Assimilative Photospheric Flux Transport (ADAPT) maps of the photospheric magnetic field topology \citep{Arge2010}, which accounts for perturbations in the solar magnetic field at specific times. For the synoptic map corresponding to the time of each eclipse observation, we use the {\fontfamily{pcr}\selectfont pfsspy} python package to generate the PFSS models \citep{Stansby2020}.
\par
PFSS models assume that the coronal plasma satisfies the time-independent, current-free, solution to Maxwell's equations (Equation \ref{eq1}) for the magnetic field. 

\begin{equation}
   \boldsymbol{\nabla} \cdot \mathbf{B} = 0
   \quad \mathrm{and} \quad
   \boldsymbol{\nabla} \times \mathbf{B} = 0
\label{eq1}
\end{equation}

Thus, for the PFSS approach, Green's function is a solution to the Laplace equation (see \citealt{Sakurai1982}; Equation \ref{eq2}).

\begin{equation}
   \mathbf{B} = - \boldsymbol{\nabla} \Phi
   \quad \mathrm{and} \quad
   \nabla^2 \Phi = 0
\label{eq2}
\end{equation}

This leads to the PFSS Solution, which is often expressed in terms of spherical harmonics (see \citealt{Wang1992}).

For this work, we first use the commonly adopted approach to define a fixed $R_{ss}$ distance. We generate PFSS models corresponding to the TSE dates between 2008-2020 for $R_{ss} = 1.3$, $1.5$, $2.0$, $2.5$, $3.0$, $3.5$, and $4.0 \ R_\odot$. 
When using {\fontfamily{pcr}\selectfont pfsspy}, field lines are shown by defining a population of ``seeds" ($S$) which propagate through the modeled magnetic field. While the PFSS model generates the magnetic field for the entire volume between the photosphere and $R_{ss}$, the chosen seed distances act as a visualization tool to represent part of the coronal magnetic field. Here we generate seeds at a set of selected radial and longitudinal distances between the photosphere and $R_{ss}$. 

While the common convention when displaying PFSS-generated field lines is to use a single seed distance right above the photosphere, we generate four sets of seeds for each PFSS model, namely at $S = 1 R_{\odot} + 0.05 R_{\odot}$, $S = 1 R_{\odot} + R_{ss}/2$, $S = 1 R_{\odot} + 2R_{ss}/3$, $S = R_{ss} - 0.01 R_{\odot}$. Thus, we initialize seed populations right above the photosphere, halfway between the photosphere and $R_{ss}$, two-thirds the way to the $R_{ss}$, and directly below the $R_{ss}$. In doing so, a large amount of both open and closed field lines are displayed. For each seed distance, we choose to generate 80 field lines equally spaced in longitude in the plane-of-sky and process each set of seeds separately. This amount gives us adequate information without having field lines too close or overlapping with each other such that they become hard to distinguish through the RHT processing (see Section \ref{ss.RHT}). In {\fontfamily{pcr}\selectfont pfsspy}, the PFSS solution is calculated on a 3D strumfric grid. The cells in the grid are defined using standard spherical coordinates $(\phi, s, \rho)$ where $\rho = ln(r)$ and $s = cos(\theta)$. For all of the PFSS simulations in this work, the number of bins in each dimension are $N_\phi=360$, $N_s=180$, and $N_\rho= 25$.

\begin{figure*}
    \centering
    \includegraphics[width=0.85\textwidth]{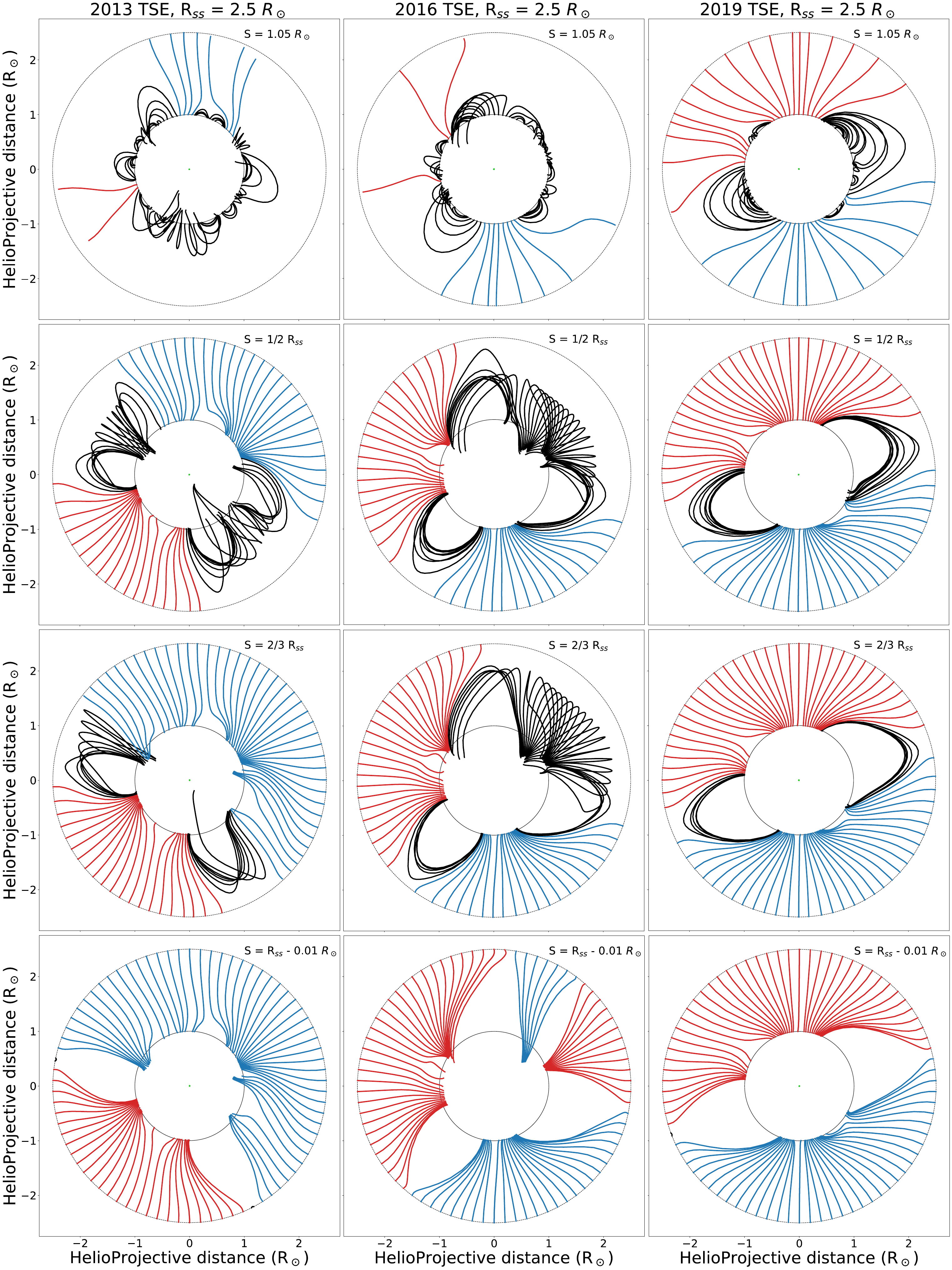}
    \caption{
    Examples of field line maps generated by the PFSS model for $R_{ss} = 2.5  \ R_\odot$, with four different seed locations as indicated in each panel.  The three columns correspond to the 2013 (solar maximum), 2016 (descending phase), and 2019 (solar minimum) eclipse dates. The same seed locations are given in each row for each year. 
   The two polarity `open' fields are given in red and blue. The black lines delineate closed field structures.
    All images are scaled to the same size with north pointed vertically up.
    }
    \label{fig.PFSS}
\end{figure*}

The combination of all the seed populations yields a much more comprehensive representation of coronal magnetic field lines, as shown in the examples of different 1-single seed choices in Figure \ref{fig.PFSS} for $R_{ss} =  2.5 R_{\odot}$. The four sequential rows show each seed population. The three columns correspond to TSE 2013 (solar maximum), 2016 (descending phase), and 2019 (solar minimum). Each population of seeds at a given heliocentric distance yields a unique representation of the magnetic field, which is why multiple seed surfaces are used here. For example, a single seed distance alone would leave significant regions of empty space in the resulting PFSS field maps, and thus would not be a complete representation of the coronal magnetic field as seen in the TSE data. It is important to note that changes in the seed distance will affect not only which field lines will be traced, but also their distribution with latitude and longitude. Albeit, small offsets in the seed locations (i.e., a few degrees) will not affect our final results as field lines near to each other roughly follow the same path.

\section{Comparison between TSE Images and PFSS Models}
\label{ss.The PFSS Model}

\subsection{Qualitative Comparison}
\label{ss.Qualitative Comparison}

\begin{figure*}
    \centering
    \includegraphics[width=0.85\textwidth]{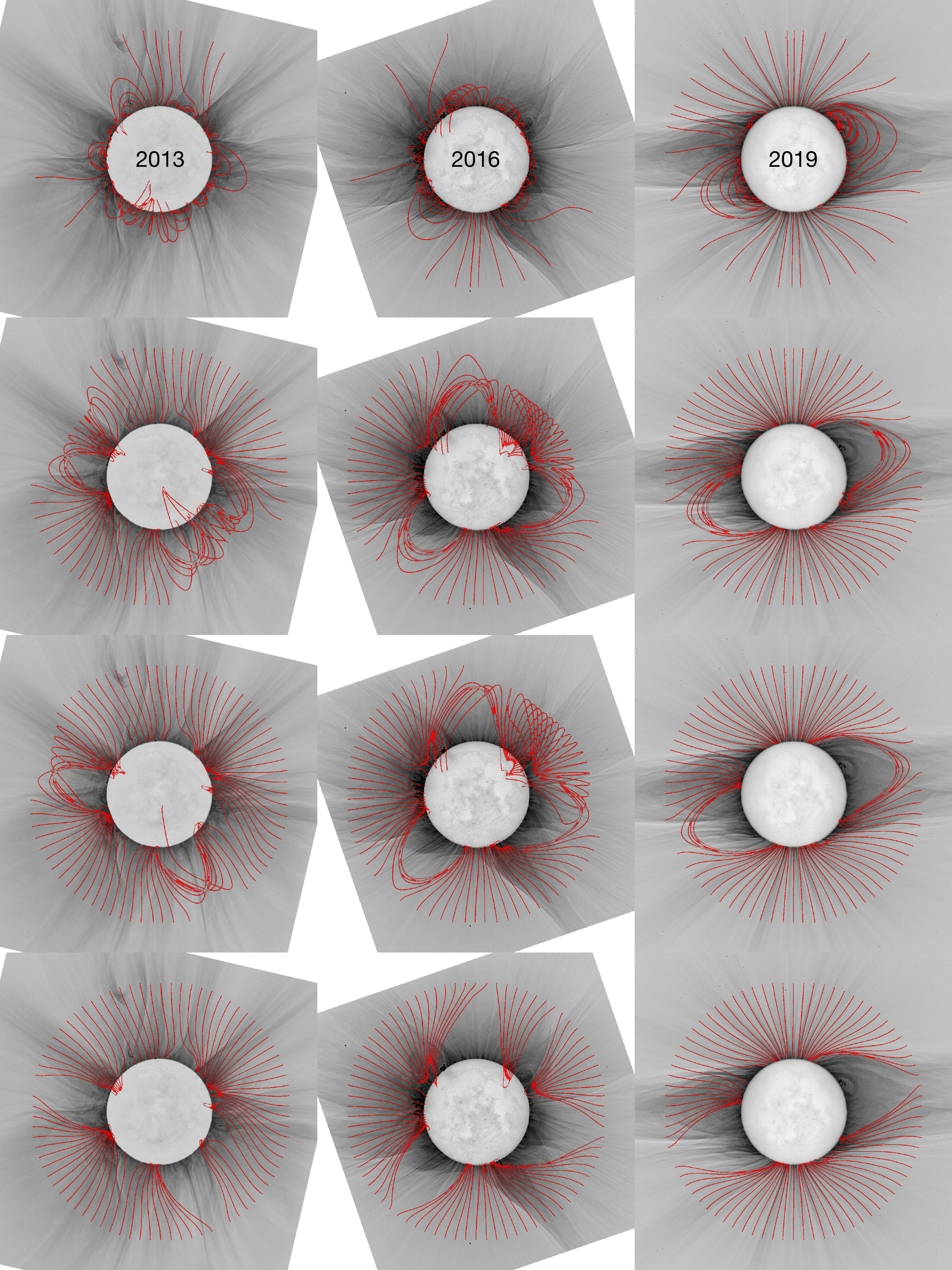}
    \caption{
    Overlay of the PFSS-generated field lines given in Figure \ref{fig.PFSS}, for $R_{ss}$ = 2.5 $ \ R_\odot$ and 4 different seed distances, over the corresponding white light TSE images for 2013, 2016, and 2019. All images are scaled to the same size with north pointed vertically up. The images are given in inverted colors with black being the highest intensity of emission.
    }
    \label{fig.PFSSoverWL}
\end{figure*}

\begin{figure*}
    \centering
    \includegraphics[width=1\textwidth]{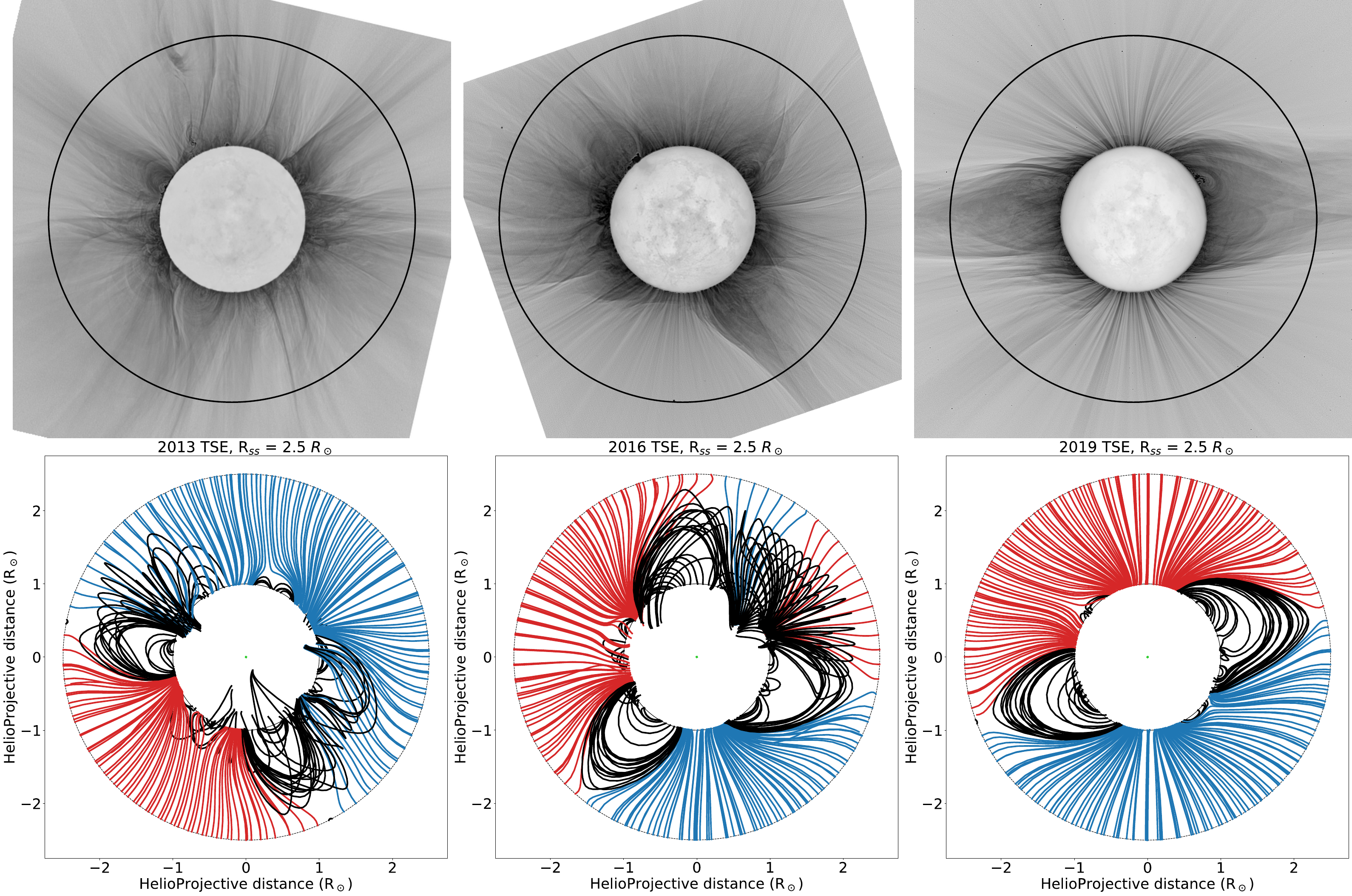}
    \caption{
    Top row: TSE white light eclipse images for 2013, 2016 and 2019. The black circle in the TSE images is at $R_{ss}$ = 2.5 $R_{\odot}$. Bottom row: PFSS-generated field lines for $R_{ss}$ = 2.5 $ \ R_\odot$ with the combination of all four seed distances, shown separately in Figure \ref{fig.PFSS}. 
    }
    \label{fig.PFSS_WL}
\end{figure*}

An overlay of the PFSS-generated magnetic field lines for $R_{ss}$ = 2.5 $  R_\odot$ and for each of the four different seed distances (Figure \ref{fig.PFSS}) on the corresponding eclipse images are shown in Figure \ref{fig.PFSSoverWL} for comparison. We find that a more comprehensive representation can be achieved by combining all four seed distances, as shown in Figure \ref{fig.PFSS_WL} for the same model and dates as shown in Figures \ref{fig.PFSS} and \ref{fig.PFSSoverWL}. There are some similar features between the TSE images and PFSS models, but there are also significant discrepancies. The closest representation of the coronal field lines seems to occur at solar minimum (2019), but the tilt angle of the streamers relative to the ecliptic plane, as evidenced in the eclipse images, is far from satisfactory.

The comparison with the TSE images demonstrates the persistent shortcomings of the $R_{ss} = 2.5 \ R_\odot$ PFSS approach. For example, at the peak of solar maximum in 2013, there are no large scale loops in the corona as generated by the PFSS. The same applies to 2016, at the descending phase, even though there are several streamers observed in the TSE images, none of them appear in the PFSS. For 2019, at solar minimum, the large scale streamers are mostly in the ecliptic plane in the TSE image but rotationally offset by at least $30^o$ in the PFSS model.

\subsection{Quantitative Comparison}
\label{ss.RHT}

To assess the departure of the PFSS modeled magnetic field lines from those depicted in TSE white light images in a more quantitative manner, we measure the angle difference between the corresponding field lines relative to the solar radial direction. To calculate this angle we first apply the 
Rolling Hough Transform (RHT) to both PFSS models and TSE images. The RHT is a modification of the Hough Transform (HT), introduced in a patent by \cite{Hough1962} for the detection of complex patterns in bubble chamber photographs. It is a machine vision algorithm that measures linear intensity as a function of orientation in images. We can use the RHT to determine the probability that each pixel in the image is part of a coherent linear structure, and the angle of the structure in the image, thus enabling the user to quantify the linearity of different structures in the plane of the image without specifying discrete entities.  

We use the {\fontfamily{pcr}\selectfont rht} python package developed by \citet{Clark2014}, which operates on two-dimensional binary images. Hence, the PFSS images are first transformed into binary images where each pixel has a value of either 1 or 0. A specified window size is then selected with which the algorithm sweeps a given image to determine the probability that any given pixel is part of a coherent linear structure. Namely, a distinct, continuous, line or curve within the given window defined in the RHT algorithm. We chose a $7\times7$ pixel window size since it is the largest pixel area which contains no more than one linear structure for the PFSS models. If we apply the RHT to the combined seed models, such as in Figure \ref{fig.PFSS_WL}, the field lines are too close to each other for the algorithm to distinguish them, so we apply the RHT to each individual seed image and co-add the data back together to get the complete representation of the coronal magnetic field. (See \cite{Schad2017} and \cite{Boe2020} for detailed description and demonstration of how the RHT algorithm can be applied to solar data to extract the magnetic field direction.)

\begin{figure*}
    \centering
    \includegraphics[width=\textwidth]{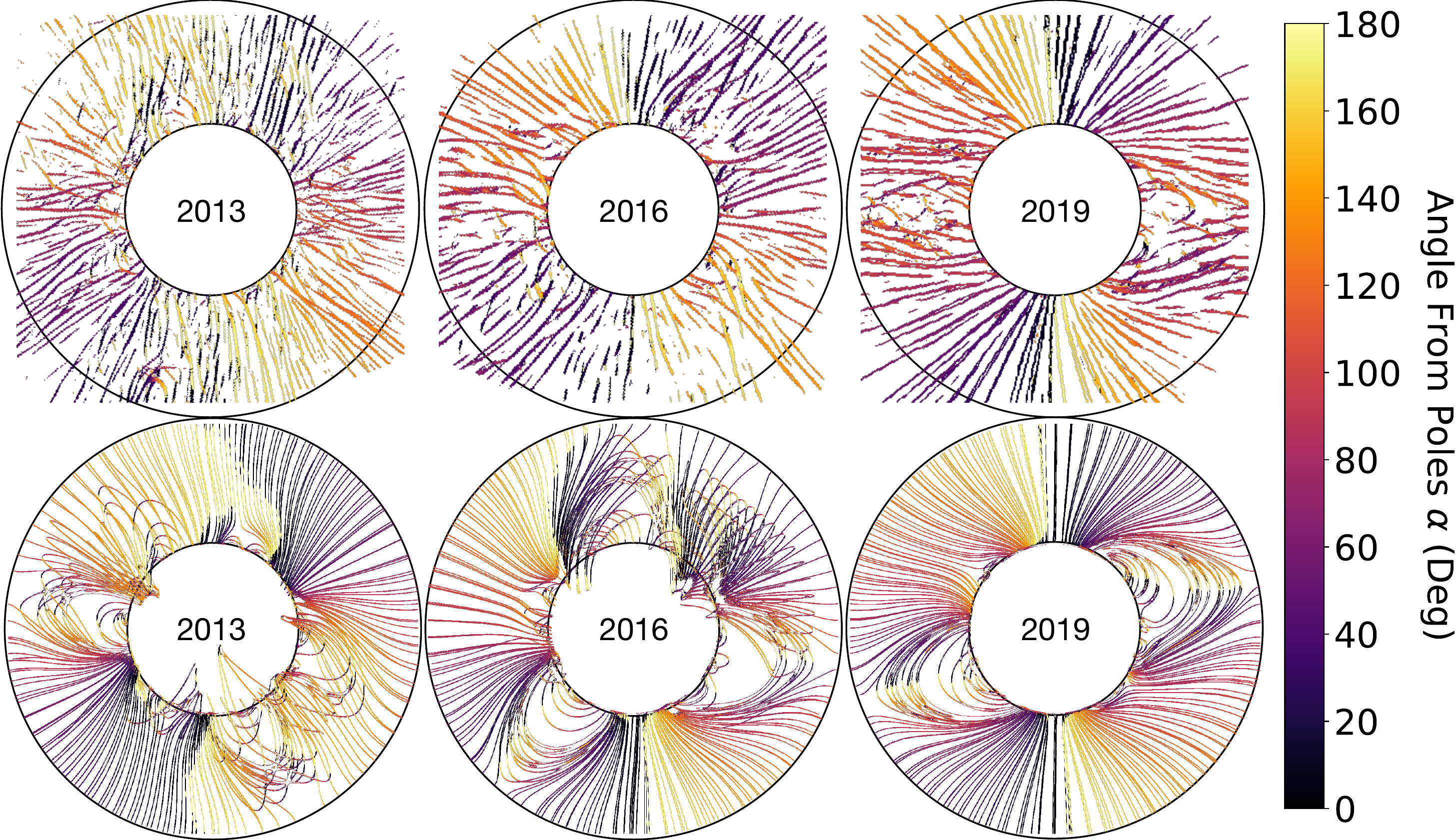}
    \caption{
    Top row: 2013, 2016, and 2019 TSE-RHT outputs with colors changing indicating change of angle relative to north and south poles. Bottom row:  the combined seeds for the PFSS-RHT $R_{ss}$ = 2.5 $ \ R_\odot$  models, with the application of the RHT. 
    All images are scaled to the same size with north pointed vertically up. (See Section \ref{ss.RHT} for details).}
    \label{fig.RHT}
\end{figure*}

The implementation of the RHT on the 2013, 2016, and 2019 TSE white light images, first presented by \cite{Boe2020}, is shown in the top row of Figure \ref{fig.RHT} and for the corresponding PFSS models from Figure \ref{fig.PFSS_WL} in the row below. The colors in the PFSS-RHT and TSE-RHT outputs represent the angles in the plane-of-sky from from $0^\circ$ to $180^\circ$, with $0^\circ$ and $180^\circ$ pointing vertically North and South and $90^\circ$ pointing equatorward. The color at $0^\circ$ is a dark purple; it changes to a lighter red as the angle with respect to north increases clockwise until it turns yellow at the south pole. The colors then flip back to purple (i.e., at $0^\circ$ relative to the south pole) and change in the same manner as they reach the north pole again (see \citet{Boe2020} for details). For closed field lines in both TSE images and PFSS models, the colors flip at the apex of the loops due to the angle change.

The RHT procedure outputs the probability that each pixel in the image has a given line direction. The most probable directions are then called angles $\alpha$. $\alpha$ is relative to $0^{\rm o}$ north or south, and the colored field lines produced by the RHT correspond to the values of $\alpha$. We also apply a sigma cut on the probability of a coherent linear structure existing on all the pixels with a certainty less than $95\%$ to get rid of the noise and leave behind only what is certain to be magnetic field lines.This is the same procedure used by \cite{Boe2020} to generate RHT outputs for the white light images done .

\begin{figure*}
    \centering
    \includegraphics[width=1\textwidth]{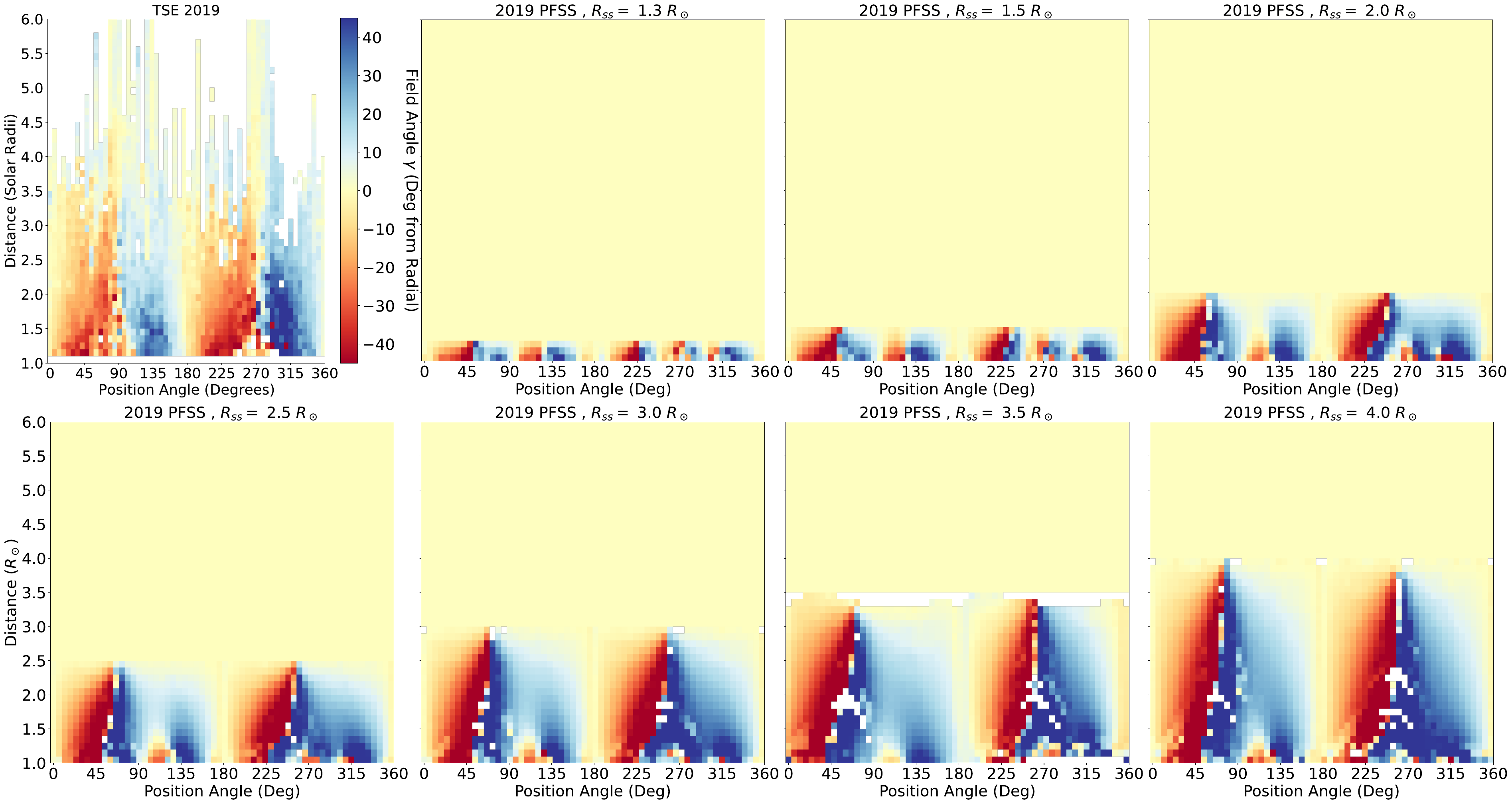}
    \caption{The top left image shows how many degrees away from being radial the coronal magnetic field is ($\gamma$) for the 2019 white light TSE image. The data is displayed as a polar coordinate histogram of distance from the Sun versus the position angle around the Sun. The rest of the images show the same histograms for the PFSS models generated with different $R_{ss}$.}
    \label{2019Track3}
\end{figure*}

The next step is to compute field angles $\gamma_{TSE}$ and $\gamma_{PFSS}$ which are the deviation of angles $\alpha$ from the radial direction relative to Sun center for both TSE structures and PSFF models (see \citealt{Boe2020}). This approach prevents any angle discontinuity effect (i.e., $0^\circ$ versus $180^\circ$) and enables a direct comparison of structures throughout the corona for each eclipse date. 
As an example, these angles are plotted versus position angle in Figure~\ref{2019Track3} for 2019. Note, position angle is measured starting from $0^\circ$ North moving counter-clockwise. The TSE `histograms' are generated with a higher pixel count than in \citet{Boe2020} to match the resolution used in the PFSS `histograms' shown in Figure~\ref{2019Track3}, where the field angles, $\gamma_{TSE}$ and $\gamma_{PFSS}$, are plotted versus position angle (horizontal axis) and radial distance (vertical axis; i.e., polar coordinates), using the color-coded vertical bar drawn between the TSE and PFSS panels. The plots are given for the seven different $R_{ss}$ values.

\begin{figure*}
    \centering
    \includegraphics[width=1\textwidth]{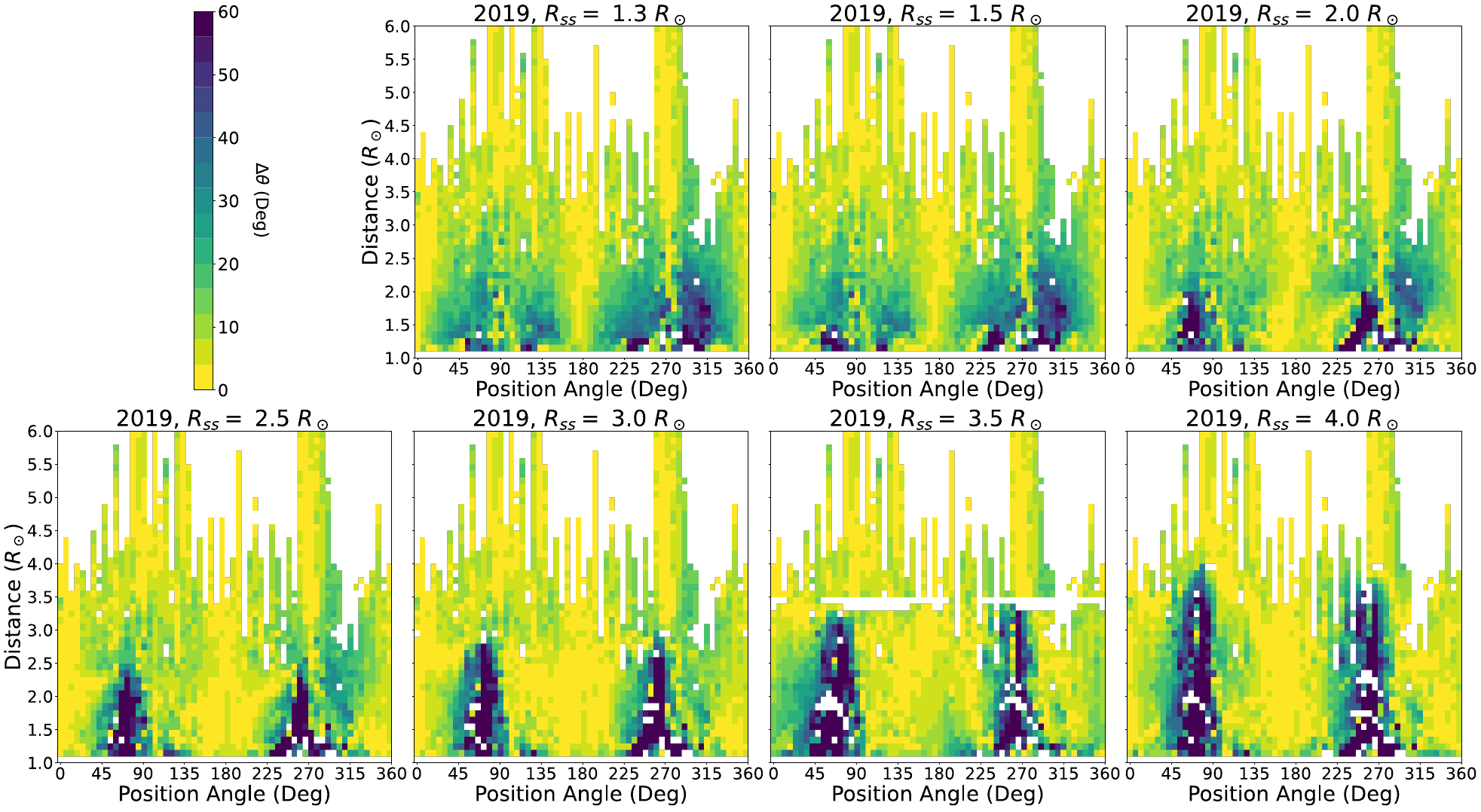}
    \caption{These histograms show $\Delta \theta$ for each different $R_{ss}$ values for the 2019 data. The histogram format is the same as Figure \ref{2019Track3}.}
    \label{2019Track4}
\end{figure*}

\begin{figure*}
    \centering
    \includegraphics[width=1\textwidth]{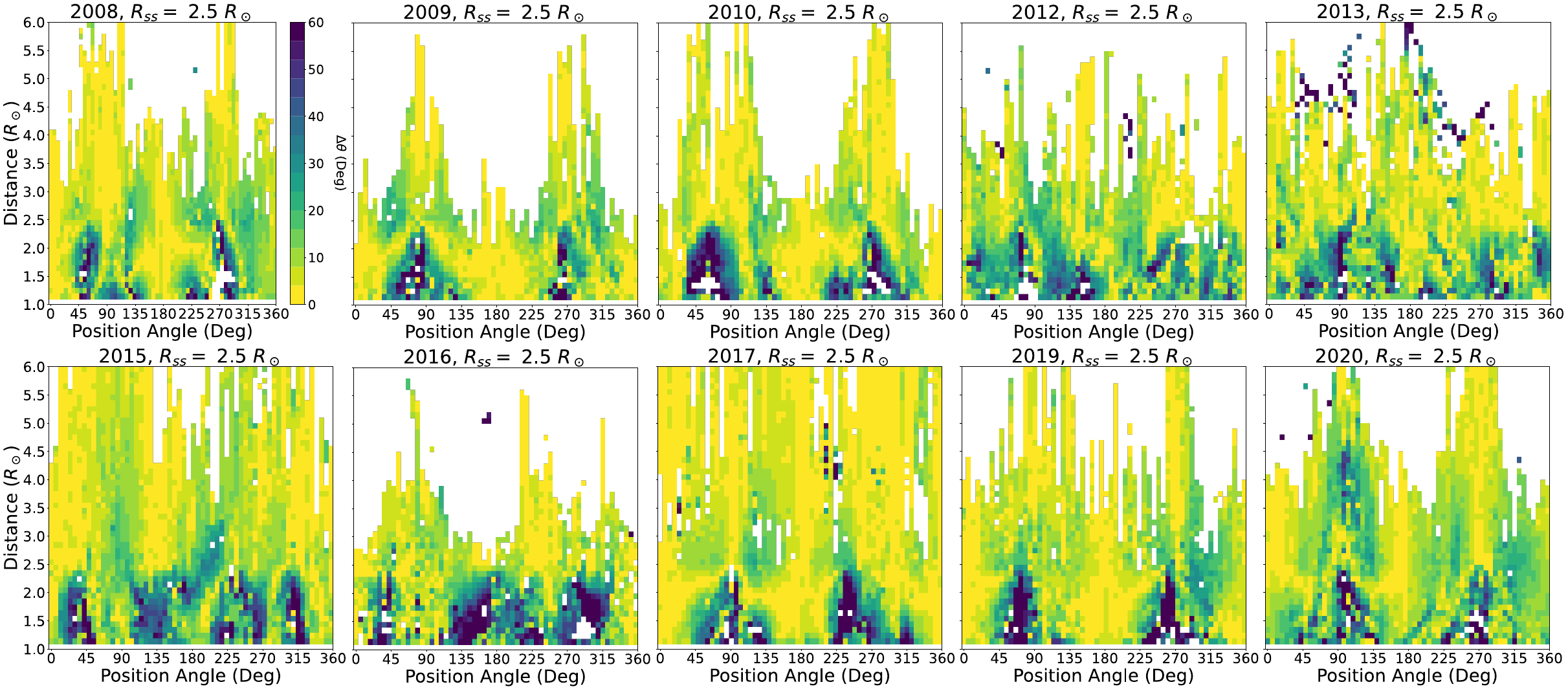}
    \caption{These histograms show $\Delta \theta$ for $R_{ss} = 2.5 \ R_\odot$ for each TSE considered. The histogram format is the same as Figure \ref{2019Track3}.}
    \label{2019Track5}
\end{figure*}

We then take the absolute value, $\Delta \theta$, of the difference between $\gamma_{TSE}$ and $\gamma_{PFSS}$ for each pixel, and we produce the $\Delta \theta$ histograms as shown in Figure~\ref{2019Track4}. The same histograms for all eclipse years for $R_{ss} = 2.5 R_\odot$ are shown in Figure~\ref{2019Track5}. 
It is important to note that the coronal mass ejections in the 2013 TSE image had strong non-radial structures at high helioprojective distances. The front of these CMEs (i.e., classic ice-cream cone shape; see \citealt{Alzate2017}) caused regions of unexpectedly large $\Delta \theta$ in Figure~\ref{2019Track5}. These regions do not have a significant impact on the findings (see Section \ref{s.Summary and Conclusions}).

\begin{figure*}
    \centering
    \includegraphics[width=1\textwidth]{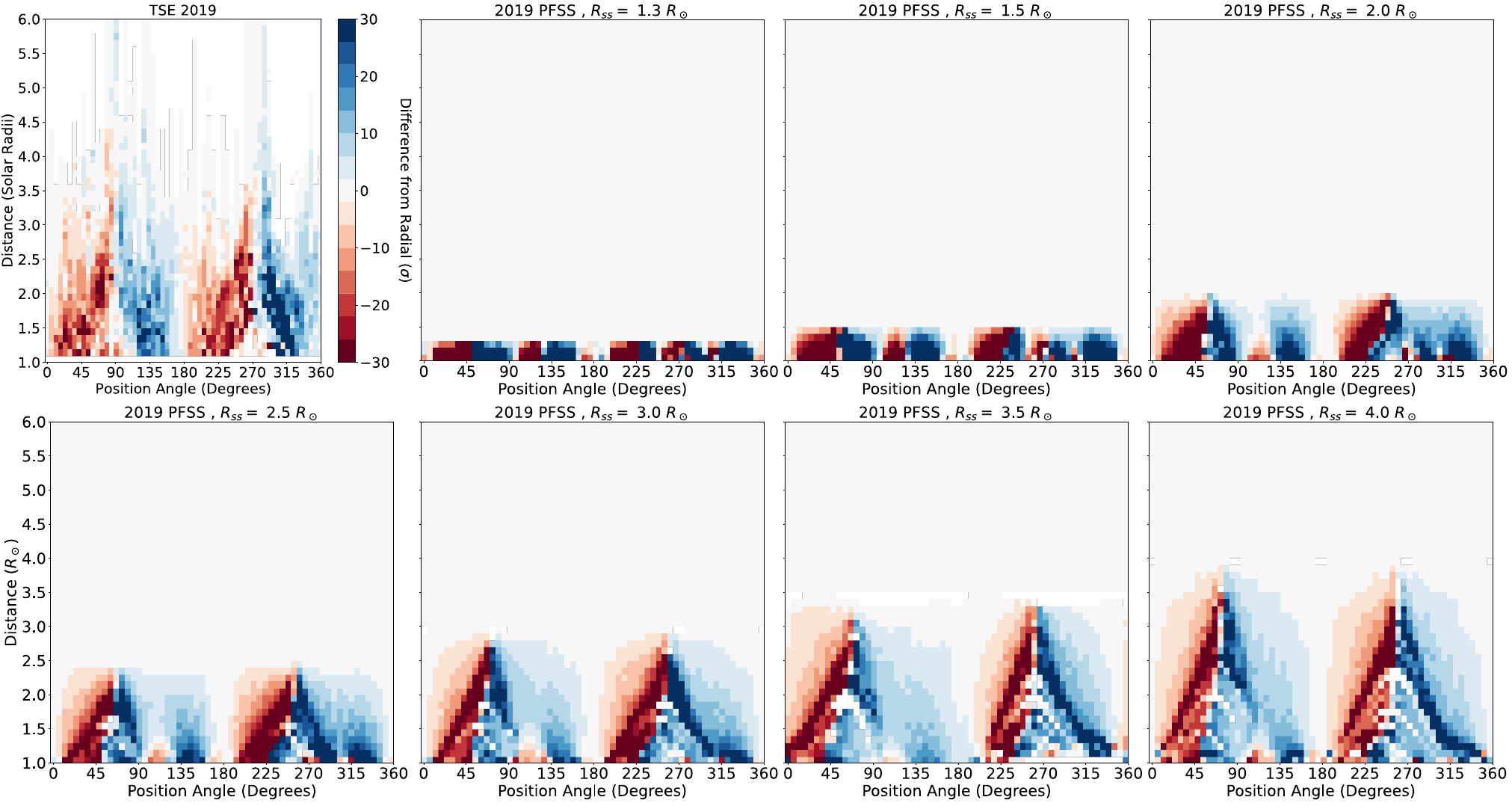} 
    \caption{The top left image shows the confidence that the magnetic field angle is not radial for the 2019 TSE ($\beta$). The histogram format is the same as Figure \ref{2019Track3}. The rest of the images shows the same histograms for the PFSS models generated with different $R_{ss}$ for the same year.}
    \label{fig.2019PFSSerrors}
\end{figure*}

\begin{figure*}
    \centering
    \includegraphics[width=1\textwidth]{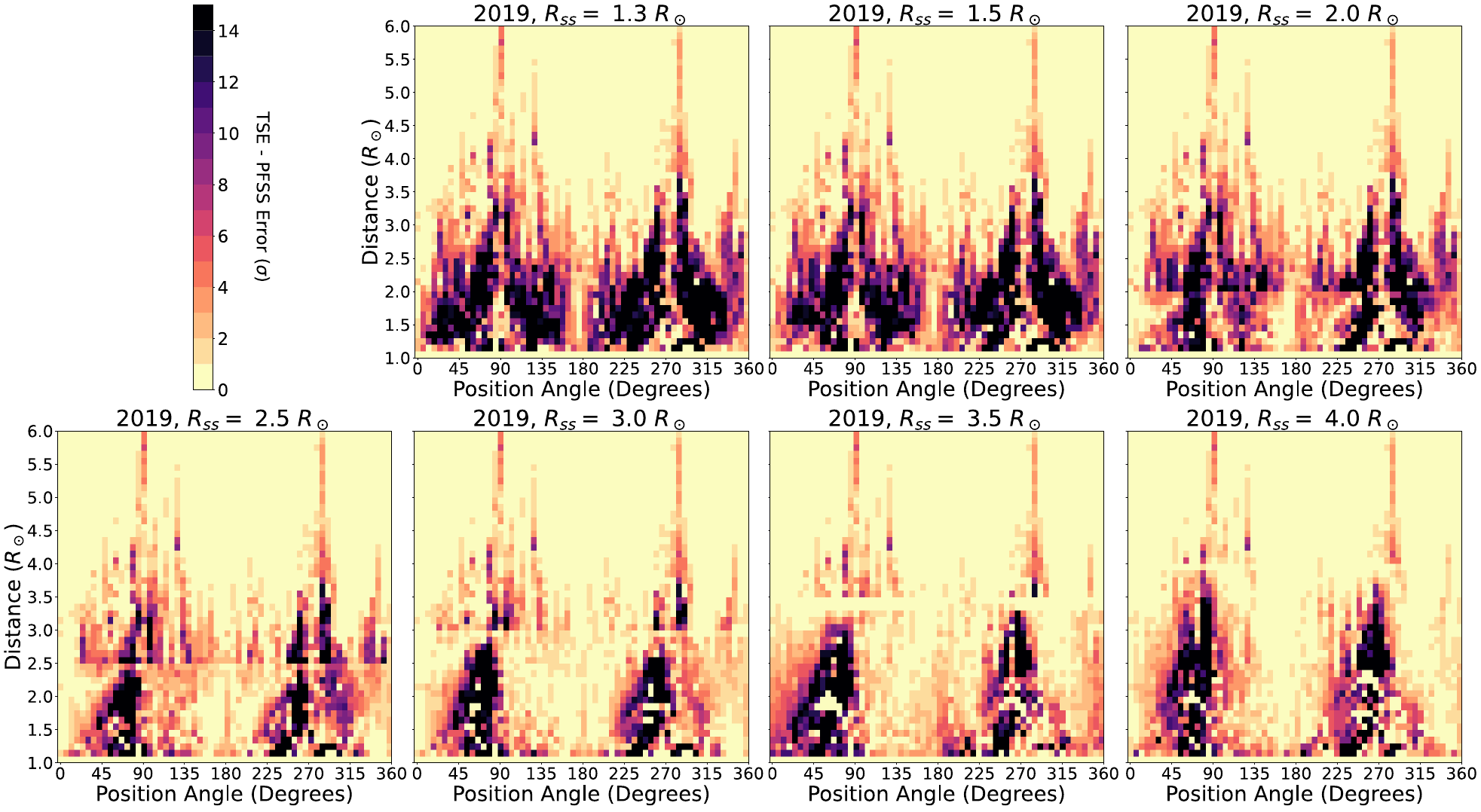} 
    \caption{These histograms show the significance of $\Delta \theta$ between the 2019 TSE data and the PFSS model for each $R_{ss}$. The histrogram format is the same as Figure \ref{2019Track3}.}
    \label{fig.2019errors}
\end{figure*}

\begin{figure*}
    \centering
    \includegraphics[width=1\textwidth]{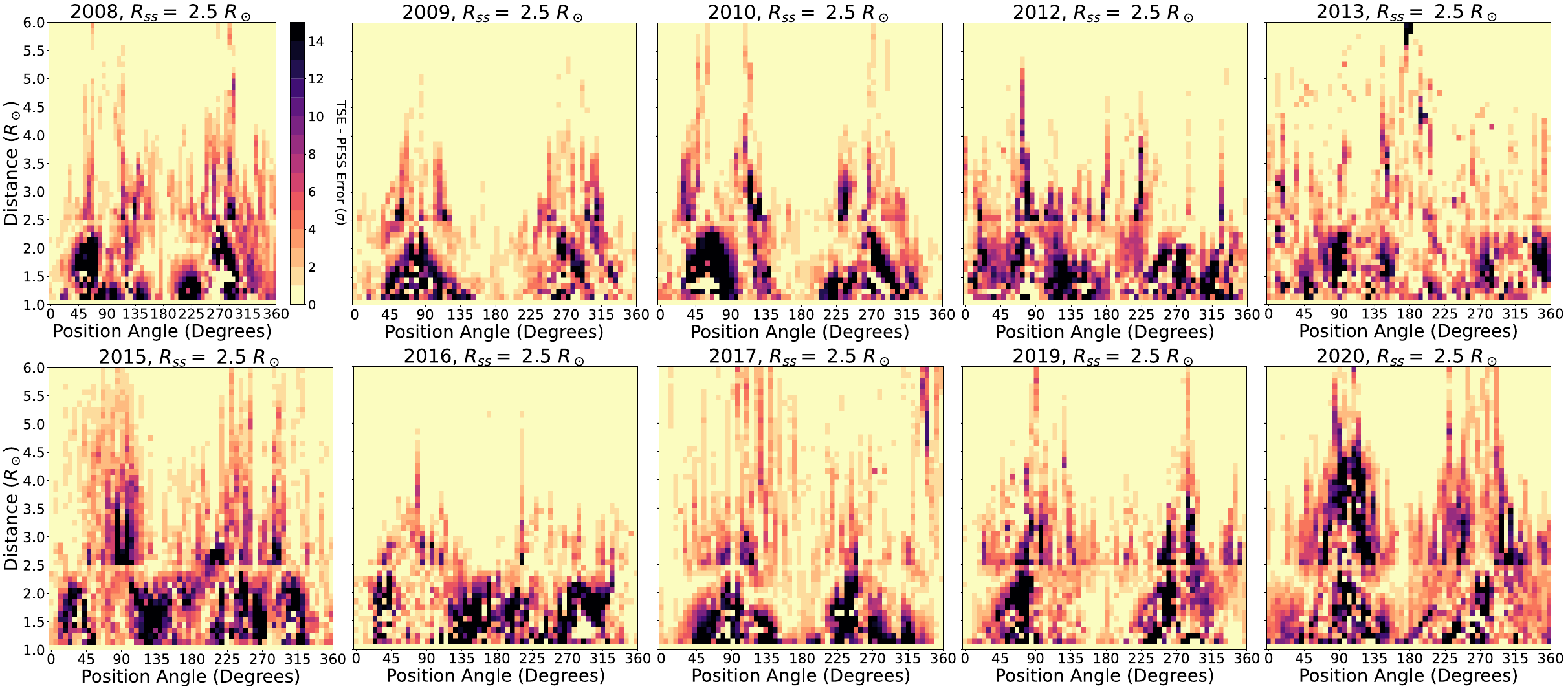} 
    \caption{These histograms show the significance of $\Delta \theta$ between the TSE data and PFSS model for each year given $R_{ss} = 2.5 \ R_\odot$. The histrogram format is the same as Figure \ref{2019Track3}.}
    \label{fig.2_5errors}
\end{figure*}

We also create plots of the significance of the coronal magnetic field being non-radial (i.e., $\gamma > 0$) using the RHT reported uncertainties,  similar to Figures \ref{2019Track3}, \ref{2019Track4}, and \ref{2019Track5}. These are shown for the 2019 data and corresponding PFSS models in Figure \ref{fig.2019PFSSerrors}. To determine the confidence that $\Delta \theta > 0$, we incorporate the uncertainty of each measurement, namely the uncertainty from the PFSS model and white light RHT process, by adding both in quadrature. 
We then show the $\Delta \theta$ significance data in Figures \ref{fig.2019errors} and \ref{fig.2_5errors} similarly to how we plotted our histograms.

\section{Discussion}
\label{s.Discussion}

The detailed comparisons between the TSE data and PFSS models reveal persistent differences. The PFSS model is often different in the magnetic field direction by greater than $10^\circ$ and sometimes as much as $60^\circ$, particularly at the bases of equatorial streamers. These differences are present for every model tested regardless of the selection of source surface. One possible explanation for these differences could be that we are not using the same part of the coronal field in both the model and eclipse (i.e., line-of-sight effects). To test our assumption that the TSE data will preferentially observe the K Corona directly in the plane-of-sky, we repeat the procedure using field lines that were generated for $\pm15^\circ$ of the plane-of-sky. Using these alternative field lines yield results that are effectively the same, such that the average $\Delta\theta$, $\langle\Delta\theta\rangle$, would only differ $\sim4^\circ$ as a maximum difference -- which is less than the measurement uncertainty and is not sufficient to explain the discrepancies between the PFSS models and TSE data (see Appendix \ref{s.Appendix}).

\pagebreak

\subsection{Solar Cycle Effects}
\label{ss.Solar Cycle Effects}

\begin{figure*}
    \centering
    \includegraphics[width=1\textwidth]{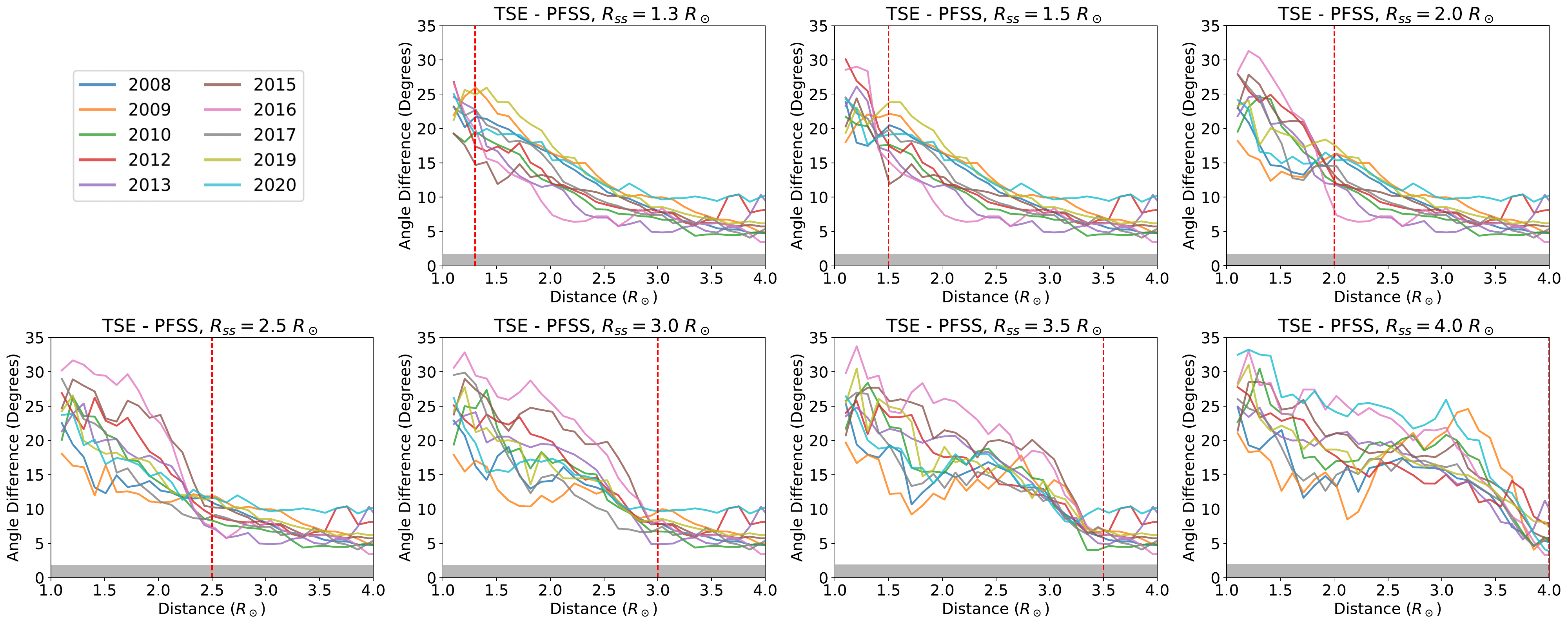} 
    \caption{These figures show $\Delta \theta$ between the PFSS models and the white light images' field line structure as a function of distance away from the photosphere, with differences being averaged over all solar latitudes at each given distance. These plots are shown for all years at each $R_{ss}$. The gray bar at the bottom represents the average confidence threshold for the data, so $\Delta \theta$ values below that threshold are consistent with no difference.}
    \label{fig.longave}
\end{figure*}

\begin{figure*}
    \centering
    \includegraphics[width=1\textwidth]{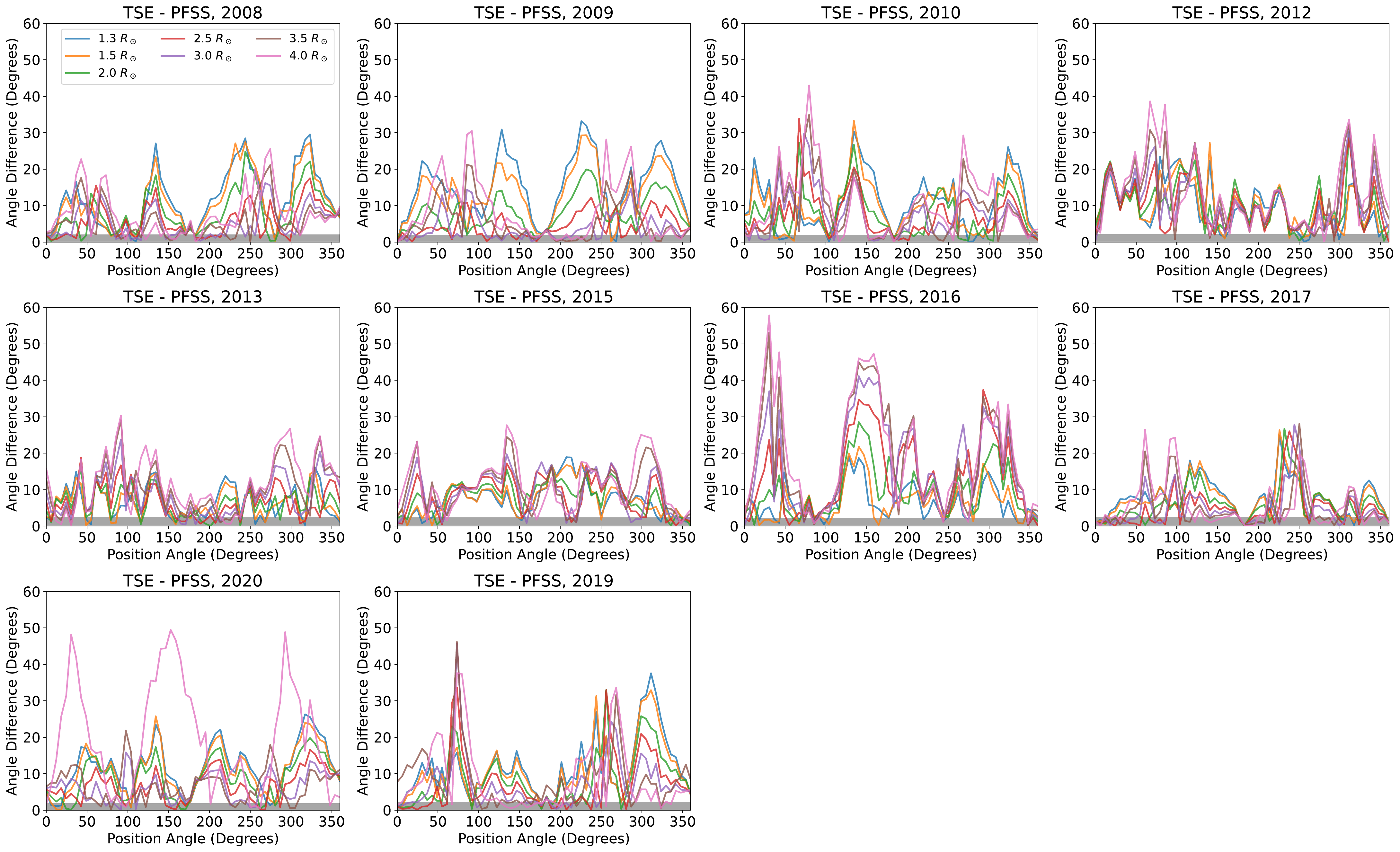} 
    \caption{These figures show $\Delta \theta$ between the PFSS models and the white light images' field line structure as a function of solar latitude, with differences being averaged over all distances away from the Sun at each solar latitude. These plots are shown for all $R_{ss}$ for each year. The gray bar at the bottom represents the average confidence threshold for the data, so $\Delta \theta$ values below that threshold are consistent with no difference.}
    \label{fig.latave}
\end{figure*}

\begin{figure*}
    \centering
    \includegraphics[width=1\textwidth]{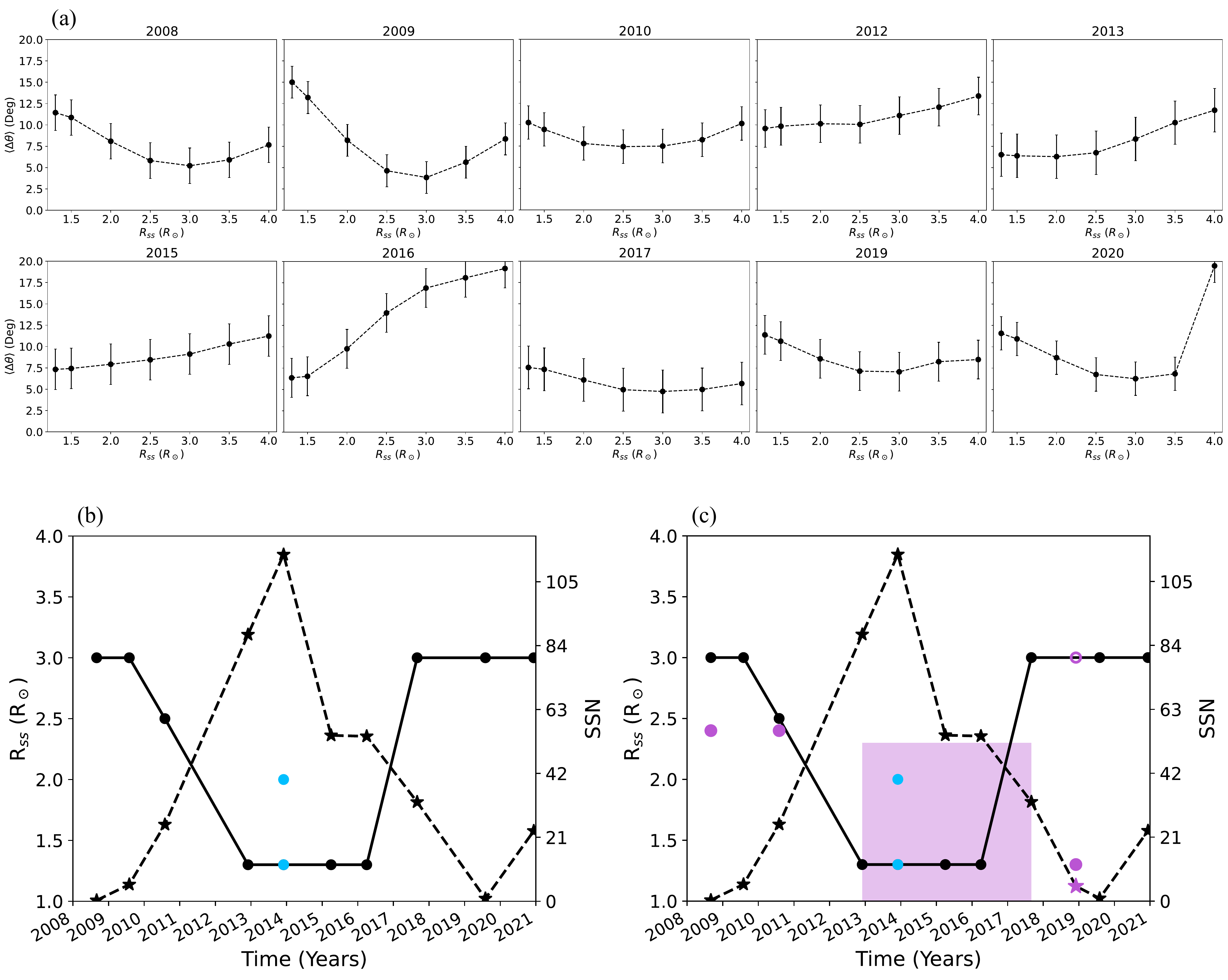} 
    \caption{
    (a) Plots of $\langle \Delta \theta \rangle$ for all TSE years as a function of $R_{ss}$.
    (b) The solid line follows the best matching $R_{ss}$ data points. The dashed line follows the sunspot numbers for the years considered. The two blue points for 2013 are the best matching values of $R_{ss} = 2.0 \ R_\odot$ and  $1.3 \ R_\odot$. 
    (c) Same as panel b, with the addition of results from \citet{Asvestari2019} (purple shaded region), \citet{Wagner2022} and \citet{Badman2020} (solid purple points).
    }
    \label{2019Track9}
\end{figure*}

Visual inspection of $\Delta \theta$ in Figures \ref{2019Track4} and \ref{2019Track5} for all the eclipse years hints at a solar cycle variation. The models seem to have a lower $\langle\Delta\theta\rangle$ for certain $R_{ss}$ values depending on what period of the solar cycle is considered. To validate the potential presence of such a trend, 
we plot $\Delta \theta$ averaged over all solar latitudes at each given distance as a function of distance away from the photosphere, as shown in Figure \ref{fig.longave}. These plots have a red dashed line to distinguish the $R_{ss}$ distance, which, beyond that line, the data only shows the observed TSE white light field angles. Additionally, we plot $\Delta \theta$ averaged over all distances away from the Sun at each solar latitude as a function of $R_{ss}$ for each year, as shown in Figure \ref{fig.latave}. In both plots, the gray shaded region on the bottom of the plot shows the confidence threshold of the measurement. That is, any data point that falls within this gray region is consistent with no difference, whereas anything outside the gray region has a statistically significant difference. Therefore, everything significantly above this gray region should be considered a bad match for the observed data.
Figure \ref{2019Track9}a shows plots of $\langle \Delta \theta \rangle$ versus $R_{ss}$ for all the dates studied. The minimum value of $\langle \Delta \theta \rangle$ should correspond to the optimal choice of $R_{ss}$ in a given year, as it gives the lowest deviation of the PFSS generated lines from those in the TSE images used as a benchmark. Uncertainties from Figures \ref{fig.2019errors} and \ref{fig.2_5errors} are incorporated into Figure \ref{2019Track9}a. Figure \ref{2019Track9}b shows the $R_{ss}$ corresponding to $\langle \Delta \theta \rangle_{min}$, i.e.,  the minimum value of $\langle \Delta \theta \rangle$, as a solid line for each year. The dashed line is the corresponding sunspot number, SSN. 

\begin{figure*}
    \centering
    \includegraphics[width=1\textwidth]{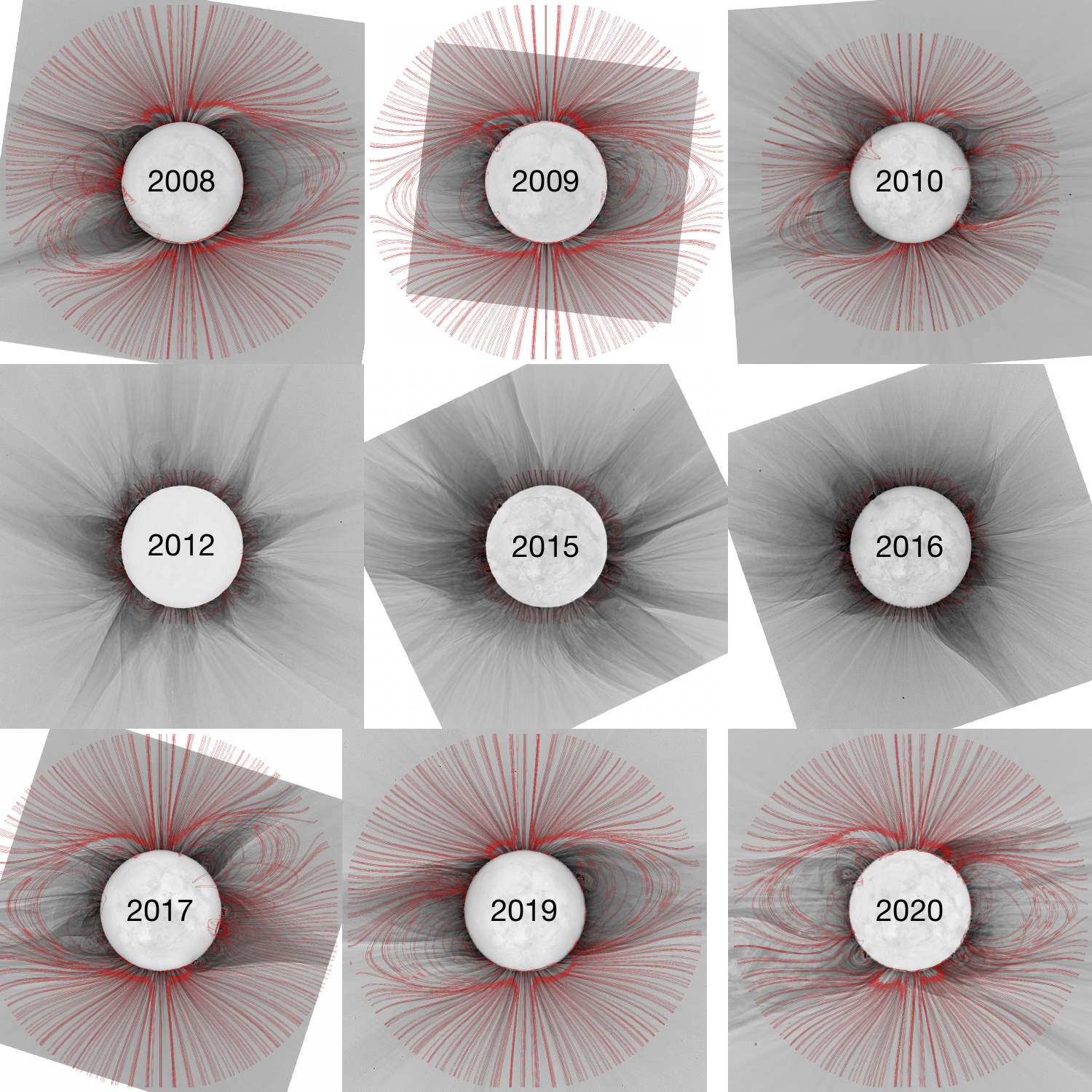}
    \caption{PFSS models for the best matching $R_{ss}$ (see Section~\ref{ss.Solar Cycle Effects}) derived from Figure \ref{2019Track9}b (red lines) overlayed on the corresponding TSE images shown out to $3 \ R_\odot$. TSE 2013 is shown separately in Figure \ref{2019Track11}.}
    \label{2019Track10}
\end{figure*}

\begin{figure*}
    \centering
    \includegraphics[width=1\textwidth]{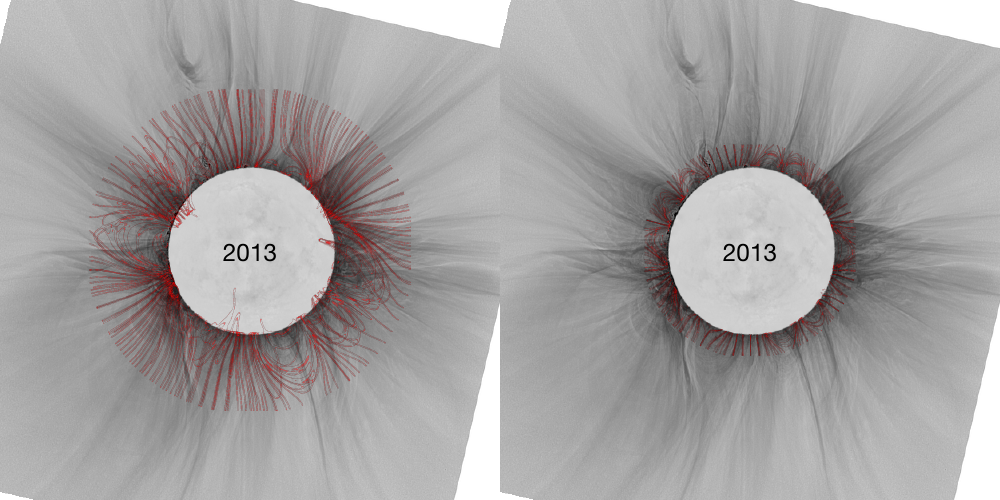}
    \caption{Same as Figure \ref{2019Track10} for the 2013 TSE in particular. This eclipse had two optimal matches to the PFSS models (See Section~\ref{ss.Special Case of 2013}) with $R_{ss}= 2.0 R_\odot$ (left) and $R_{ss}=1.3R_\odot$ (right), shown here overlayed on the corresponding TSE image.}
    \label{2019Track11}
\end{figure*}

By using TSE images to benchmark PFSS models, we find that there is a clear anti-correlation between $\langle \Delta \theta \rangle_{min}$ and SSN with $R_{ss}$ increasing to 3.0 around solar minimum, and decreasing to 1.3 around solar maximum. However, variations in solar activity can occur during the intermediary time between eclipses which can impact the best fitting $R_{ss}$. Hence, we cannot unequivocally conclude that the $R_{ss}$ depends entirely on the solar cycle. Regardless of potential change, these examples spread throughout the solar cycle indicate a consistent trend. To acquire a better assessment of the validity of the optimal $R_{ss}$ values thus inferred, we show an overlay of the corresponding `optimal' PFSS-generated field lines over the corresponding TSE images in Figure \ref{2019Track10}. The somewhat abnormal 2013 year is shown separately in Figure \ref{2019Track11} for the two optimal values of $R_{ss}$ = 1.3 and 2.0. Note that in both figures, the PFSS modeled field lines cannot exceed $R_{ss}$ in distance. It is clear from Figs. \ref{2019Track10} and \ref{2019Track11} that the optimal PFSS models match the TSE images a lot better than the first examples of Figs. \ref{fig.PFSSoverWL} and \ref{fig.PFSS_WL}.

\subsection{Special Case of 2013}
\label{ss.Special Case of 2013}

One exception to the expected trend in Figure \ref{2019Track9}b is 2013, where there are two possible minima for $\langle \Delta \theta \rangle$, at $R_{ss}$ = 1.2 and 2.0, as shown in blue points. For the 2013 PFSS model, the best matching $R_{ss}$ increases when the expected trend is to stay at 1.3 $ \ R_\odot$. When looking at $\langle \Delta \theta \rangle$ for all the $R_{ss}$ for the 2013 model, the difference between the 1.3 $ \ R_\odot$ and 2.0$ \ R_\odot$ varies only by $\sim 0.4^{\rm o}$ from the minimum $\langle \Delta \theta \rangle$ at 2.0 $ \ R_\odot$. To verify this finding, we repeated our procedure with ADAPT synoptic maps $\pm$ 1 day and $\pm$ 1 solar rotation ($\sim$ 1 month) from the TSE dates used. This data was then run through the same analysis process as explained above to determine the best fitting $R_{ss}$. It was found that the best match of $R_{ss}$ was invariant for $\pm$ 1 day and $\pm$ 1 month for all the TSE dates we used except for the 2013 data. 

For the 2013 TSE date, taking $\pm$ 1 day would maintain a best match of $R_{ss}$ = 2.0 $  R_\odot$. Further exploration of different rotations changes the best matching $R_{ss}$ to 1.5 $  R_\odot$ for October and September, and it would stay at 2.0 $  R_\odot$ through December and the month after. We note that the 2013 TSE captured a CME in the southwest quadrant of the corona \citep{Alzate2017} that could have altered the coronal structure. Consequently, we applied the process again,  leaving out the CME at position angles between $90^\circ$ to $180^\circ$. No change to the best matching $R_{ss}$ = 2.0 $  R_\odot$ was found, implying that the best match is 2.0 $  R_\odot$, but marginally so when compared to 1.3 $  R_\odot$ and 1.5 $  R_\odot$.

\pagebreak

\subsection{Comparison with Other Models}
\label{ss.Comparison with Other Models}

We now compare our results with recent studies by \citet{Asvestari2019}, \citet{Badman2020}, and \citet{Wagner2022}. Their data points are shown in Figure \ref{2019Track9}c. \citet{Asvestari2019} used synoptic maps of EUV observations from 2012 to 2017. They found that $ R_{ss} \leq 2.3 ~ R_\odot$ yields a best match for the coronal hole boundaries. Their ranges are shown in the purple shaded region in Figure \ref{2019Track9}c. This is in relatively good agreement with our results with the exception of 2017 when our optimal value was 3.0 $  R_\odot$. On the other hand, \citet{Badman2020} considered the magnetic field measurements taken by the FIELDS on board the Parker Solar Probe in November 2018 for their study to trace field lines back to small scale polarity inversions at the photosphere. Their best value of $R_{ss} = 1.3 \ R_\odot$ is shown as the solid purple point in Figure \ref{2019Track9}c. In contrast, when running our calculation for that date, which is the closest to our 2019 TSE, we find that the best matching value is $R_{ss}$ = 3.0 $ \ R_\odot$. Lastly, \citet{Wagner2022} looked at observational data from both the Solar and Heliospheric Observatory/Large Angle and Spectrometric COronagraph (SOHO/LASCO) and enhanced solar eclipse photographs produced by Druckmüller and the Solar TErrestrial RElations Observatory/Sun Earth Connection Coronal and Heliospheric Investigation (STEREO/SECCHI) for the 2008 and 2010 TSE dates. The coronal modeling they compare with is taken from the EUropean Heliospheric FORecasting Information Asset (EUHFORIA), and the ``coronal domain" of this model uses PFSS and SCS. They conclude that $R_{ss} = 2.4 ~ R_\odot$ matched their observations the best which we find is similar to our 2010 best matching $R_{ss}$ but not with 2008.

Finally, we point out that \citet{Asvestari2019} had attempted to investigate a solar cycle dependent $R_{ss}$, as originally proposed by \citet{Lee2011} and \citet{Arden2014}. However, the \citet{Asvestari2019}'s data set, which included coronal hole boundaries, did not cover a full cycle and they were not able to detect any solar cycle trend. \citet{Asvestari2020} performed a follow-up study using the same sample of coronal holes, and they concluded that the performance of their model was unrelated to coronal hole characteristics and the solar cycle.
\citet{Lee2011} looked at solar cycles 22 and 23, and they found that $1.5 \ R_\odot \leq R_{ss} \leq 1.9 \ R_\odot$ was an optimal range of $R_{ss}$ values depending on the solar cycle.
\citet{Arden2014} looked at solar cycles 23 and 24. They concluded that that an $2.875 \ R_\odot \leq R_{ss} \leq 3.25 \ R_\odot$ were optimal values for solar minimum, and that $R_{ss} \sim 2.5 \ R_\odot$ was optimal for solar maximum. 
Although \citet{Lee2011} and \citet{Arden2014} both explored a solar cycle dependant $R_{ss}$ value, we find that the ranges of optimal  $R_{ss}$ values should be larger than what they concluded.

\section{Summary and Conclusions}
\label{s.Summary and Conclusions}

This study evaluates the reliability of PFSS generated field lines by using a complete solar cycle of total solar eclipse white light images acquired between 2008 and 2020 as a benchmark. These are the only data available at present which can provide high spatial resolution traces of coronal magnetic field in the most relevant region of the corona, namely from the solar surface up to at least 6 $ \ R_\odot$, where the corona undergoes its most complex evolution. 

Comparison of PFSS generated field lines and TSE images reveals that the conventional value for the source surface of $R_{ss}$ = 2.5 $ \ R_\odot$ yields significant differences often exceeding $10^o$, sometimes more than $60^o$, with a relatively poor match for the location and structure of streamers (as shown in Figure~\ref{fig.PFSS_WL}). 

One interesting outcome of this work is the finding that the optimal $R_{ss}$ values, for this sample of TSEs, seem to be solar cycle dependent with the lowest at 1.3 $ \ R_\odot$ for solar maximum, increasing to 3.0 $ \ R_\odot$ at solar minimum. As illustrated in Figs. \ref{2019Track10} and \ref{2019Track11}, the match between PFSS and TSE is greatly improved by using this variable $R_{ss}$. Still, significant differences persist between the PFSS models and the eclipse data for all iterations of the models.

It is clear from the TSE images in Figs. \ref{2019Track10} and \ref{2019Track11}, that the corona within $3 ~ R_\odot$ at and close to solar maximum is dominated by low lying loops. Indeed, the ambiguous case of 2013 discussed earlier in association with Figure \ref{2019Track9}, can be resolved through the comparison of the two options for $R_{ss}$ = 2 and 1.3 $ \ R_\odot$, shown in the left and right panels respectively in Figure \ref{2019Track11}. The value of 1.3 $ \ R_\odot$ matches remarkably well the low-lying closed structures. In contrast, the value of 2.0 $ \ R_\odot$ presents PFSS field lines that extend further out in the corona which introduce spurious streamers and yield a poor match to the fields lines in the TSE image. 

It is worth noting that TSE images show that open field lines are just as prevalent near solar maximum as solar minimum regardless of the presence of active regions and streamers. As for closed coronal structures,  Figure \ref{2019Track10} shows how their sizes increase significantly to produce broader streamers near solar minimum, compared to shorter and smaller closed field regions near solar maximum. 

Despite certain $R_{ss}$ values minimizing the differences between the PFSS models and TSE data, our findings indicate that that PFSS models generally do a poor job of replicating the coronal magnetic field regardless of the choice of $R_{ss}$. Further, the traditional $R_{ss}$ of $\sim  2.5 \ R_\odot$ often performs worse than other choices. This optimization is particularly critical for tracing solar wind streams measured in-situ back to their sources at the Sun and to establish the location of the heliospheric current sheet. Implementation of these optimal values is equally important for MHD models which use PFSS models as a starting boundary condition. 

Our findings strongly suggest that PFSS models alone are unreliable for detailed tracing of the origin of the solar wind. The often large differences of the magnetic field structure could lead to significant errors in any such analysis. The inferred and actual sources of the solar wind may be quite different when there are consistently 10-30$^\circ$ degree differences between the PFSS models and TSE white light images. 

More detailed MHD simulations are able to do noticeably better at matching the coronal magnetic field and even the electron temperature and density (e.g., \citealt{Boe2021a,Boe2022,Boe2023}), but require substantially more computational power which is not as easily available for large scale studies. Thus, PFSS models should be used with extreme caution when attempting to map the precise sources of the solar wind through the corona. However, they are a useful tool for approximating the coronal magnetic field when less angular precision is required.

To their credit, it is quite remarkable that PFSS models can even come close to producing complex coronal structures, especially at solar maximum. Further, PFSS models do considerably better in the low corona (below $\sim 1.5 \ R_\odot$). In particular, in strong magnetic field regions. Thus, PFSS models are still useful for mapping active regions and prominence cavities, and for generating initial conditions for other models. Indeed, the issues with the PFSS simulations could largely be explained by changes in the plasma $\beta$ as a function of height. Thus, PFSS simulations work best when plasma $\beta$ is near 0 (e.g., around active regions and prominence cavities when the magnetic pressure is much larger than the gas pressure).
Perhaps the differences between PFSS and MHD models could be used to illuminate the effect of plasma $\beta$ and ways that computationally cheap PFSS simulations could be improved in the future.

\section*{Acknowledgements}
Support for this work was provided by NSF grants AST-2303171 and AGS-2313853 to the Institute for Astronomy of the University of Hawai`i-Mānoa. Financial support was provided to B. B. by the National Science Foundation under Award No. 2028173. This work utilizes data produced collaboratively between Air Force Research Laboratory (AFRL) \& the National Solar Observatory (NSO). The ADAPT model development is supported by AFRL. The input data utilized by ADAPT is obtained by NSO/NISP (NSO Integrated Synoptic Program). NSO is operated by the Association of Universities for Research in Astronomy (AURA), Inc., under a cooperative agreement with the National Science Foundation (NSF). The ADAPT maps were obtained at \href{https://gong.nso.edu/adapt/maps}{https://gong.nso.edu/adapt/maps}.


\pagebreak

\appendix
\section{Ruling out Line-of-Sight Effects}
\label{s.Appendix}

One possible explanation for the significant differences between the PFSS models and TSE data could be line-of-sight (LOS) effects (see Section \ref{s.Discussion}). In particular, if any field lines observed during the eclipse are sufficiently out of the plane-of-sky (POS), the projected angle could be somewhat different from the actual magnetic field. The procedure we used in this work involved generating the PFSS lines on the POS at the initialization point, though the field lines would often move in or out of the POS as they propagated through the PFSS magnetic field. 
\par
\begin{figure*}[h!]
    \centering
    \includegraphics[width=0.9\textwidth]{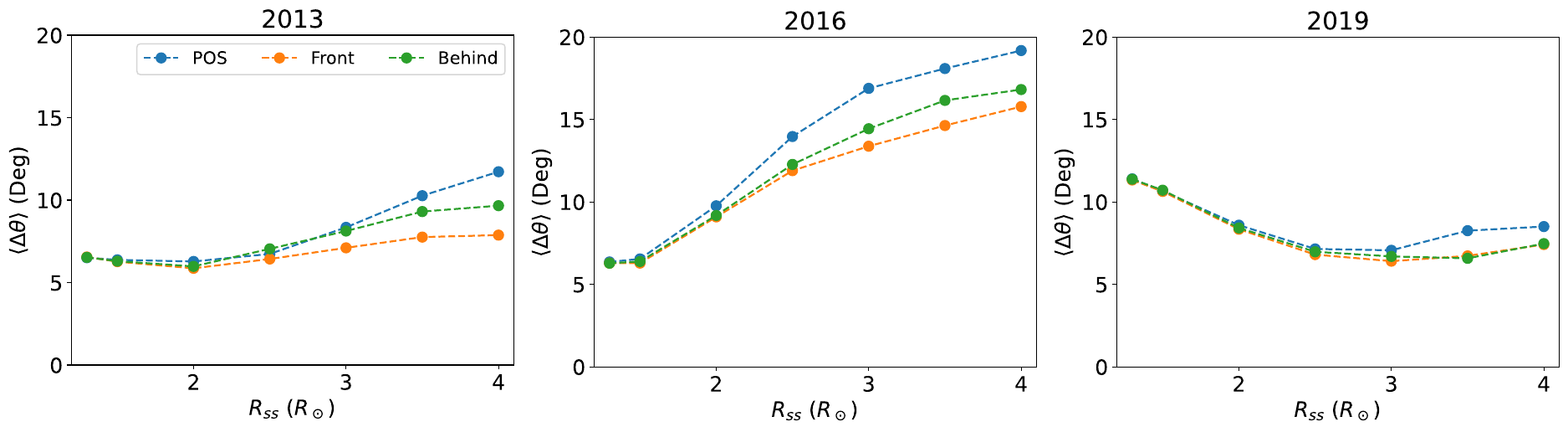} 
    \caption{The $\langle \Delta \theta \rangle$ between the PFSS models and TSE data for the set of all tested source surfaces (same as Figure \ref{2019Track9}a). For each of the three eclipses shown here, field lines were generated 15$^\circ$ in front of and behind the POS to test LOS effects (see Appendix \ref{s.Appendix}).}
    \label{Apen2}
\end{figure*}

To test whether any LOS-related effect could interfere with our findings, we repeated the PFSS and TSE comparison using field lines generated $\pm 15^\circ$ longitude of the POS (i.e., one set in front, one behind) for the 2013, 2016, and 2019 TSEs. We ran these new test cases for all the source surface values used throughout the work. The average difference between the PFSS model and TSE data, $\langle \Delta \theta \rangle$, for each test case are shown in Figure~\ref{Apen2}. Considering a wider range of longitudes did not yield significant changes in the average $\langle \Delta \theta \rangle$, with only small changes of a few degrees for source surfaces beyond 3 $R_\odot$. Indeed, all cases were virtually identical for lower source surfaces. Since the maximum changes are about equal to the measurement uncertainty, and are much smaller than the overall difference between the data and model, this test indicates that effects related to the POS are not responsible for the difference between the PFSS models and TSE data.
\par
Further, by observing the projected image of the PFSS field lines, rather than their true angle in the model, we are comparing an observable that is analogous to the TSE observation. By design, this procedure should suppress any discrepancies between the PFSS prediction and TSE data.

\pagebreak

\bibliographystyle{apj}
\bibliography{main_v2}

\begin{thebibliography}{}
\expandafter\ifx\csname natexlab\endcsname\relax\def\natexlab#1{#1}\fi

\bibitem[{{Altschuler} \& {Newkirk}(1969)}]{Altschuler1969}
{Altschuler}, M.~D., \& {Newkirk}, G. 1969, \solphys, 9, 131

\bibitem[{{Alzate} {et~al.}(2017){Alzate}, {Habbal}, {Druckm{\"u}ller}, {Emmanouilidis}, \& {Morgan}}]{Alzate2017}
{Alzate}, N., {Habbal}, S.~R., {Druckm{\"u}ller}, M., {Emmanouilidis}, C., \& {Morgan}, H. 2017, \apj, 848, 84

\bibitem[{{Arden} {et~al.}(2014){Arden}, {Norton}, \& {Sun}}]{Arden2014}
{Arden}, W.~M., {Norton}, A.~A., \& {Sun}, X. 2014, Journal of Geophysical Research (Space Physics), 119, 1476

\bibitem[{{Arge} {et~al.}(2010){Arge}, {Henney}, {Koller}, {Compeau}, {Young}, {MacKenzie}, {Fay}, \& {Harvey}}]{Arge2010}
{Arge}, C.~N., {Henney}, C.~J., {Koller}, J., {et~al.} 2010, in American Institute of Physics Conference Series, Vol. 1216, Twelfth International Solar Wind Conference, ed. M.~{Maksimovic}, K.~{Issautier}, N.~{Meyer-Vernet}, M.~{Moncuquet}, \& F.~{Pantellini}, 343--346

\bibitem[{{Asvestari} {et~al.}(2019){Asvestari}, {Heinemann}, {Temmer}, {Pomoell}, {Kilpua}, {Magdalenic}, \& {Poedts}}]{Asvestari2019}
{Asvestari}, E., {Heinemann}, S.~G., {Temmer}, M., {et~al.} 2019, Journal of Geophysical Research (Space Physics), 124, 8280

\bibitem[{Asvestari {et~al.}(2020)Asvestari, Heinemann, Temmer, Pomoell, Kilpua, Magdalenic, \& Poedts}]{Asvestari2020}
Asvestari, E., Heinemann, S.~G., Temmer, M., {et~al.} 2020, Journal of Physics: Conference Series, 1548, 012004

\bibitem[{Badman {et~al.}(2020)Badman, Bale, Oliveros, Panasenco, Velli, Stansby, Buitrago-Casas, Réville, Bonnell, Case, de~Wit, Goetz, Harvey, Kasper, Korreck, Larson, Livi, MacDowall, Malaspina, Pulupa, Stevens, \& Whittlesey}]{Badman2020}
Badman, S.~T., Bale, S.~D., Oliveros, J. C.~M., {et~al.} 2020, The Astrophysical Journal Supplement Series, 246, 23

\bibitem[{{Boe} {et~al.}(2023){Boe}, {Downs}, \& {Habbal}}]{Boe2023}
{Boe}, B., {Downs}, C., \& {Habbal}, S. 2023, \apj, 951, 55

\bibitem[{{Boe} {et~al.}(2021{\natexlab{a}}){Boe}, {Habbal}, {Downs}, \& {Druckm{\"u}ller}}]{Boe2021a}
{Boe}, B., {Habbal}, S., {Downs}, C., \& {Druckm{\"u}ller}, M. 2021{\natexlab{a}}, \apj, 912, 44

\bibitem[{Boe {et~al.}(2022)Boe, {Habbal}, {Downs}, \& {Druckm{\"u}ller}}]{Boe2022}
Boe, B., {Habbal}, S., {Downs}, C., \& {Druckm{\"u}ller}, M. 2022, \apj, 935, 173

\bibitem[{{Boe} {et~al.}(2020){Boe}, {Habbal}, \& {Druckm{\"u}ller}}]{Boe2020}
{Boe}, B., {Habbal}, S., \& {Druckm{\"u}ller}, M. 2020, \apj, 895, 123

\bibitem[{{Boe} {et~al.}(2021{\natexlab{b}}){Boe}, {Yamashiro}, {Druckm{\"u}ller}, \& {Habbal}}]{Boe2021b}
{Boe}, B., {Yamashiro}, B., {Druckm{\"u}ller}, M., \& {Habbal}, S. 2021{\natexlab{b}}, \apjl, 914, L39

\bibitem[{Clark {et~al.}(2014)Clark, Peek, \& Putman}]{Clark2014}
Clark, S.~E., Peek, J. E.~G., \& Putman, M.~E. 2014, The Astrophysical Journal, 789, 82

\bibitem[{{Darwin} {et~al.}(1899){Darwin}, {Schuster}, \& {Maunder}}]{Darwin1899}
{Darwin}, L., {Schuster}, A., \& {Maunder}, E.~W. 1899, RSPTA

\bibitem[{{Druckm{\"u}ller}(2009)}]{Druckmuller2009}
{Druckm{\"u}ller}, M. 2009, The Astrophysical Journal, 706, 1605

\bibitem[{{Druckm{\"u}ller} {et~al.}(2006){Druckm{\"u}ller}, {Ru{\v{s}}in}, \& {Minarovjech}}]{Druckmuller2006}
{Druckm{\"u}ller}, M., {Ru{\v{s}}in}, V., \& {Minarovjech}, M. 2006, Contributions of the Astronomical Observatory Skalnate Pleso, 36, 131

\bibitem[{{Habbal} {et~al.}(2021){Habbal}, {Druckm{\"u}ller}, {Alzate}, {Ding}, {Johnson}, {Starha}, {Hoderova}, {Boe}, {Constantinou}, \& {Arndt}}]{Habbal2021}
{Habbal}, S.~R., {Druckm{\"u}ller}, M., {Alzate}, N., {et~al.} 2021, \apjl, 911, L4

\bibitem[{{Hale}(1908)}]{Hale1908}
{Hale}, G.~E. 1908, \apj, 28, 315

\bibitem[{Hough(1962)}]{Hough1962}
Hough, P.~V. 1962

\bibitem[{{Howard} \& {Tappin}(2009)}]{Howard2009}
{Howard}, T.~A., \& {Tappin}, S.~J. 2009, \ssr, 147, 31

\bibitem[{{Lee} {et~al.}(2011){Lee}, {Luhmann}, {Hoeksema}, {Sun}, {Arge}, \& {de Pater}}]{Lee2011}
{Lee}, C.~O., {Luhmann}, J.~G., {Hoeksema}, J.~T., {et~al.} 2011, \solphys, 269, 367

\bibitem[{{Levine} {et~al.}(1982){Levine}, {Schulz}, \& {Frazier}}]{Levine1982}
{Levine}, R.~H., {Schulz}, M., \& {Frazier}, E.~N. 1982, \solphys, 77, 363

\bibitem[{{Linker} {et~al.}(1999){Linker}, {Miki{\'c}}, {Biesecker}, {Forsyth}, {Gibson}, {Lazarus}, {Lecinski}, {Riley}, {Szabo}, \& {Thompson}}]{Linker1999}
{Linker}, J.~A., {Miki{\'c}}, Z., {Biesecker}, D.~A., {et~al.} 1999, \jgr, 104, 9809

\bibitem[{{Maguire} {et~al.}(2020){Maguire}, {Carley}, {McCauley}, \& {Gallagher}}]{Maguire2020}
{Maguire}, C.~A., {Carley}, E.~P., {McCauley}, J., \& {Gallagher}, P.~T. 2020, \aap, 633, A56

\bibitem[{{Maunder}(1899)}]{Maunder1988}
{Maunder}, E.~W. 1899, {The Indian eclipse, 1898 : report of the expeditions organized by the British Astronomical Association to observe the total solar eclipse of 1898 January 22} (London: Hazell, Watson, and Viney)

\bibitem[{{Miki{\'c}} {et~al.}(1999){Miki{\'c}}, {Linker}, {Schnack}, {Lionello}, \& {Tarditi}}]{Mikic1999}
{Miki{\'c}}, Z., {Linker}, J.~A., {Schnack}, D.~D., {Lionello}, R., \& {Tarditi}, A. 1999, Physics of Plasmas, 6, 2217

\bibitem[{{Munro} \& {Withbroe}(1972)}]{Munro1972}
{Munro}, R.~H., \& {Withbroe}, G.~L. 1972, \apj, 176, 511

\bibitem[{{Newkirk}(1967)}]{Newkirk1967}
{Newkirk}, Gordon, J. 1967, Annual Review of Astronomy and Astrophysics, 5, 213

\bibitem[{{Saito}(1958)}]{Saito1958}
{Saito}, K. 1958, \pasj, 10, 49

\bibitem[{{Sakurai}(1979)}]{Sakurai1979}
{Sakurai}, T. 1979, \pasj, 31, 209

\bibitem[{Sakurai(1981)}]{Sakurai1981}
Sakurai, T. 1981, \solphys, 69, 343

\bibitem[{{Sakurai}(1982)}]{Sakurai1982}
{Sakurai}, T. 1982, \solphys, 76, 301

\bibitem[{{Schad}(2017)}]{Schad2017}
{Schad}, T. 2017, \solphys, 292, 132

\bibitem[{{Schatten} {et~al.}(1969){Schatten}, {Wilcox}, \& {Ness}}]{Schatten1969}
{Schatten}, K.~H., {Wilcox}, J.~M., \& {Ness}, N.~F. 1969, \solphys, 6, 442

\bibitem[{{Schulz} {et~al.}(1978){Schulz}, {Frazier}, \& {Boucher}}]{Schulz1978}
{Schulz}, M., {Frazier}, E.~N., \& {Boucher}, D.~J., J. 1978, \solphys, 60, 83

\bibitem[{{Schwabe}(1844)}]{Schwabe1844}
{Schwabe}, H. 1844, Astronomische Nachrichten, 21, 233

\bibitem[{Song(2023)}]{Song2023}
Song, Y.-C. 2023, Research in Astronomy and Astrophysics, 23, 075020

\bibitem[{Stansby {et~al.}(2020)Stansby, Yeates, \& Badman}]{Stansby2020}
Stansby, D., Yeates, A., \& Badman, S.~T. 2020, Journal of Open Source Software, 5, 2732

\bibitem[{{van der Holst} {et~al.}(2014){van der Holst}, {Sokolov}, {Meng}, {Jin}, {Manchester}, {T{\'o}th}, \& {Gombosi}}]{vanDerHolst2014}
{van der Holst}, B., {Sokolov}, I.~V., {Meng}, X., {et~al.} 2014, \apj, 782, 81

\bibitem[{{Wagner} {et~al.}(2022){Wagner}, {Asvestari}, {Temmer}, {Heinemann}, \& {Pomoell}}]{Wagner2022}
{Wagner}, A., {Asvestari}, E., {Temmer}, M., {Heinemann}, S.~G., \& {Pomoell}, J. 2022, \aap, 657, A117

\bibitem[{{Wang} \& {Sheeley}(1992)}]{Wang1992}
{Wang}, Y.~M., \& {Sheeley}, N.~R., J. 1992, \apj, 392, 310

\bibitem[{{Zhao} {et~al.}(2017){Zhao}, {Landi}, {Lepri}, {Gilbert}, {Zurbuchen}, {Fisk}, \& {Raines}}]{Zhao2017}
{Zhao}, L., {Landi}, E., {Lepri}, S.~T., {et~al.} 2017, \apj, 846, 135

\end{thebibliography}


\end{document}